\documentstyle[psfig]{mn}

\def\etal{et~al.}
\def\spose#1{\hbox to 0pt{#1\hss}}
\def\lta{\mathrel{\spose{\lower 3pt\hbox{$\mathchar"218$}}
     \raise 2.0pt\hbox{$\mathchar"13C$}}}
\def\gta{\mathrel{\spose{\lower 3pt\hbox{$\mathchar"218$}}
     \raise 2.0pt\hbox{$\mathchar"13E$}}}

\def\clean{{\sc clean}}

\def\aips{{\sc aips}}
\def\robust{{\sc robust}}
\def\jmfit{{\sc jfmit}}
\def\imagr{{\sc imagr}}

\def\kms{\,km\,s\,$^{-1}$}
\def\Ho50{$H_0 = 50$km\,s$^{-1}$\,Mpc$^{-1}$}

\title[Radio sources in MS1054$-$03]{${\mathbf \mu}$Jy radio sources in
the ${\mathbf z=0.83}$ cluster MS1054$-$03}

\author[P.~N.~Best \etal]{P.~N.~Best,$^1$\thanks{Email: pnb@roe.ac.uk}
P. G. {van Dokkum},$^2$\thanks{Hubble Fellow} M. Franx$^3$ and
H. J. A. R{\"o}ttgering$^3$ \\ $^1$ Institute for Astronomy, Royal
Observatory Edinburgh, Blackford Hill, Edinburgh EH9 3HJ, UK\\ $^2$
California Institute of Technology, Mail Stop 105-24, 1200 E California
Blvd, Pasadena, CA 91125, USA\\ $^3$ Sterrewacht Leiden, Postbus 9513,
2300RA Leiden, the Netherlands }

\pagerange{\pageref{firstpage}--\pageref{lastpage}}
\pubyear{1999}
\begin{document}
\label{firstpage}

\maketitle

\begin{abstract}
\noindent An extremely deep 5\,GHz radio observation is presented of the
rich cluster MS1054$-$03 at redshift $z=0.83$. 34 radio sources are
detected down to a 6$\sigma$ level of 32$\mu$Jy, compared to about 25
expected from previous blank field radio source count determinations; the
sources giving rise to these excess counts lie predominantly within 2
arcmins ($\sim 700$\,kpc) of the cluster centre. Existing imaging and
spectroscopic observations have provided optical identifications for 21 of
the radio sources and redshifts for 11, of which 8 are confirmed cluster
members. Four of these eight confirmed cluster sources are associated with
close galaxy pairs (10 -- 25 kpc projected offset) of similar magnitude,
implying that the radio source may be triggered by an
interaction. However, although MS1054$-$03 has a very high fraction (17\%)
of on--going mergers (separations $\ll 10$\,kpc), no radio emission is
detected towards any of these merger events, setting a mean upper limit of
10$M_{\odot}$\,yr$^{-1}$ for any star formation associated with these
mergers.  This supports a hypothesis that low luminosity radio sources may
be onset by initial weak interactions rather than direct mergers. The host
galaxies of the other four confirmed cluster radio sources are all
isolated, and show a range of morphologies from early--type to Sc. A
comparison between the emission line and radio luminosities suggests that
two of these four radio sources are low--luminosity active galactic nuclei
(AGN), whilst for at least one of the other two the radio emission is
associated with on--going star formation. All of the radio sources
associated with the galaxy pairs appear more likely AGN than starburst
origin. The overall proportion of radio sources associated with AGN in
this cluster ($\gta 75$\%) is higher than that detected at these flux
density levels in the field (40$-$50\%).
\end{abstract}


\section{Introduction}
\label{intro}

Clusters of galaxies are the largest, most massive, collapsed structures
in the Universe and, although only a small fraction ($\approx 5$\%;
Dressler 1984) \nocite{dre84} of all low redshift galaxies lie in cluster
environments, clusters play a key role in studies of galaxy evolution
since they contain large numbers of galaxies at the same distance.  The
central regions of clusters are dominated by a population of luminous
early-type galaxies which occupy a narrow locus in colour--magnitude
relations and show a tight relationship between their effective radius,
effective surface brightness, and central velocity dispersion (the
fundamental plane, Dressler \etal\ 1987; Djorgovski \& Davies
1987)\nocite{dre87a,djo87b}; these relations evolve with redshift out to
$z = 0.83$ roughly in accordance with passive evolution predictions, with
no significant increase in scatter (e.g. van Dokkum \etal\ 1998a,b,
Schade, Barrientos \& Lopez-Cruz 1997, Stanford, Eisenhardt \& Dickinson
1998).\nocite{sch97,dok98b,dok98a,sta98} However, Butcher \& Oemler
\shortcite{but78} showed that by redshift $z \sim 0.3$ a substantial
population of bluer galaxies appears in many, but not all, clusters. These
bluer galaxies are almost all disk galaxies, of which many exhibit strong
H$\alpha$ or [OII]~3727 emission lines indicative of recent star formation
(e.g. Dressler et~al. 1997, 1999).\nocite{dre97,dre99} These results point
strongly to a higher level of activity in high redshift clusters.

In addition to star--formation activity, galaxies can also display nuclear
activity associated with an active galactic nucleus (AGN). One of the
cleanest methods of searching for either of these forms of activity is to
utilise the radio emission. AGN are extremely luminous at radio
wavelengths, with luminosities reaching as high as about
$10^{28}$\,W\,Hz$^{-1}$ at 1.4\,GHz. Normal (ie non-AGN) galaxies have a
range of observed radio luminosities, between about $10^{18}$ and
$10^{24}$\,W\,Hz$^{-1}$ at 1.4\,GHz, with starbursting galaxies filling
the upper half of that range. The radio emission arising from these normal
galaxies comes predominantly from the synchrotron emission of particles
accelerated in supernova shocks (e.g. Condon 1992 and references therein),
with a smaller thermal contribution from H{\rm II} regions, and so
essentially reflects the current star formation of the galaxy.

At redshifts $z \gta 0.2$ the radio luminosities of luminous starburst
galaxies and weak AGN correspond to microjansky flux density levels, and
over recent years numerous $\mu$Jy level radio surveys of the field have
been carried out \cite{don87,fom91,win93,ric98,mux99,gar00a}.  The
integral radio source count shows an upturn below a couple of mJy,
indicating the emergence of a new population of radio sources at
microjansky levels \cite{win84,win85}; indeed, a large proportion of the
$\mu$Jy radio source population have been shown to be associated with
starburst galaxies and normal spiral galaxies at substantial redshifts $z
\gg 0.2$ (e.g. Windhorst et~al 1995)\nocite{win95}.  The steep slope of
the differential source counts of $\mu$Jy radio sources, similar to that
of the very powerful radio galaxies and quasars, implies that this
population of sources, if it is a single population, is undergoing strong
cosmological evolution with either the density or luminosity of the
sources being higher in the past. This is in broad agreement with
determinations of the evolution of the cosmic star formation rate
\cite{mad98}, suggesting that studies of $\mu$Jy radio sources are
important for understanding the star formation history and evolution of
ordinary galaxies \cite{haa00}.

Whilst deep $\mu$Jy field surveys have made considerable progress in
recent years, no studies of radio sources in distant clusters have been
carried out to comparable depth. At low redshift ($z<0.09$), the radio
luminosity function in clusters, when viewed in terms of the proportion of
optical galaxies which have a given radio luminosity, is statistically
indistinguishable from that of the field \cite{led96}. This is somewhat
unexpected since the onset of a starburst or AGN is likely to be induced
either by an infall of gas onto the galaxy or through a weak interaction
or merger with a companion galaxy: it would be natural to assume that each
of these phenomena would occur at a different rate in cluster environments
than in the field. In low redshift clusters the similarity may be due to
the relatively relaxed states of the clusters, with much of the gas and
galaxies lying in stable virialised orbits and the gas having been
stripped from the galaxies in the cluster centre (e.g. Gunn \& Gott
1972)\nocite{gun72}: these effects reduce interactions and prevent a
higher fraction of radio sources forming in these environments. Clusters
of galaxies at higher redshifts, however, may still be in their formation
process, with relatively high galaxy merger rates (e.g. van Dokkum \etal\
1999)\nocite{dok99a} and a plentiful supply of disturbed gas. These should
provide ideal laboratories to induce starbursts and AGN (cf. the high
fraction of post--starburst galaxies in $z \sim 0.5$ clusters; Poggianti
et~al 2000).\nocite{pog00}

Dwarakanath \& Owen \shortcite{dwa99} carried out a detailed radio study
of two $z \approx 0.25$ clusters, Abell 2125 and Abell 2645, which have
similar redshifts and richnesses (Abell class 4), but very different
fractions of Butcher--Oemler blue galaxies; Abell 2125 has a blue galaxy
fraction of 0.19, whereas that of Abell 2645 is only 0.03. They found that
the radio luminosity distribution of cluster members of Abell 2125 is
bimodal, with a peak at about $10^{24.5}$\,W\,Hz$^{-1}$ composed entirely
of AGN, and a second higher peak at or below (due to the detection limit)
$10^{22.5}$\,W\,Hz$^{-1}$ composed of a mixture of AGN and star-forming
galaxies (see Section~\ref{lumdist} for more details). In Abell 2645 many
fewer radio sources were detected, with essentially the entire lower
luminosity class of sources missing. The presence of this lower luminosity
class of radio sources therefore seems to be connected with the presence
of the blue Butcher--Oemler galaxies, although the two populations do not
overlap strongly: only one of the blue galaxies in the Abell 2125 cluster
has an associated radio source.

Smail et~al \shortcite{sma99b} have made a deep radio observation (rms
$1.4 \times 10^{22}$\,W\,Hz$^{-1}$ in our assumed cosmology) of the
$z=0.41$ cluster CL0939$+$4713 and detect 8 cluster radio
sources\footnote{Here the blended entries labelled 296 and 230 by Smail
et~al.\ are taken to be part of the same extended radio source, since even
if separated into two sources the far brighter 296 entry still remains
resolved.}, associated with a range of galaxy morphologies from
ellipticals through to Sd--type galaxies. At higher redshifts, Stocke \etal\
\shortcite{sto99b} have studied radio galaxies in clusters with redshifts
$0.3 < z \lta 0.8$ to a limiting point source 1.4\,GHz radio luminosity of
$10^{23.5}$\,W\,Hz$^{-1}$.  They found no evidence for evolution of the
population of radio sources between redshift $z \sim 0.8$ and the present
epoch. However, their observations are only sensitive enough to detect the
brighter of the two populations of cluster radio sources discovered by
Dwarakanath \& Owen \shortcite{dwa99}, and it is the fainter population
which appears to depend upon the dynamical state of the cluster, and hence
may be expected to show strong redshift evolution.
 
The goal of the current project, therefore, is to investigate in detail
the nature of the fainter radio source population in high redshift
clusters. This aim was addressed by carrying out a deep VLA observation of
the rich cluster MS1054$-$03 at $z=0.83$. The nature of this cluster and
the existing observations of it are described in Section~\ref{ms1054}, and
the new VLA observations are described in Section~\ref{vlaobs}. The radio
source population is investigated and compared with the optical imaging
data in Section~\ref{radpop}. In Section~\ref{hostgals} an analysis is
made of the cluster radio source population, including a detailed
comparison of the radio, optical and emission line properties of the
cluster galaxies. Conclusions are drawn in Section~\ref{concs}. Throughout
the paper, cosmological parameters of $\Omega = 1$, $\Lambda = 0$ and
$H_{0}= 70$\,km\,s$^{-1}$Mpc$^{-1}$ are assumed.

\section{The ${\mathbf z=0.83}$ cluster MS1054$-$03}
\label{ms1054}

With a redshift of $z=0.83$, MS1054$-$03 is the highest redshift cluster
in the Extended Medium Sensitivity Survey (EMSS) X--ray selected cluster
sample \cite{gio90}, and also one of the most luminous in X--rays with a
0.3--3.5\,keV X--ray luminosity of $9 \times 10^{44}$\,erg\,s$^{-1}$. On
account of its high redshift and richness (Abell class 3), a deep 5 by 5
arcminute mosaic was made of the field around the cluster using the
wide--field planetary camera 2 (WFPC2) on the Hubble Space Telescope
(HST), in two filters (F606W and F814W; van Dokkum \etal\ 2000). This has
subsequently been supplemented by deep near--infrared imaging of the
cluster in the $J$, $H$ and $K$ wavebands using the NTT, and $U$, $B$, $V$
imaging using FORS on the VLT, to provide galaxy colours across a long
wavelength baseline (Franx \etal\ in preparation). These images show a
conspicuous overdensity of red cluster galaxies with a somewhat irregular
and elongated distribution \cite{dok00}, probably consisting of three
sub-clumps of galaxies at the same radial velocity; this is consistent
with the detection of substructure in the X--ray image \cite{don98,jel01}.

Multi--object spectroscopy carried out on the Keck Telescope has enabled
redshifts to be determined for over 200 objects in this field, with more
than 130 of these being confirmed as cluster members (van Dokkum \etal\
2000, Tran \etal\ in preparation). These authors found that the fraction
of all early type galaxies in the central regions of the cluster is 46\%,
much lower than that at low redshift ($\sim 80$\%). Further, a very high
fraction (17\%) of cluster galaxies are classified as ``merger/peculiar''
on the basis of double nuclei (separations $\ll 10\,$kpc), tidal tails,
and distorted morphologies \cite{dok99a}. This high fraction strongly
argues against monolithic collapse models of galaxy formation.
Interestingly, many of the merging galaxies are red, bulge--dominated
galaxies with no detected nebular line emission, and colours offset from
the early--type $(U-B)_z$ colour magnitude relation by only 0.07
magnitudes. The fraction of blue galaxies in the cluster, calculated in a
manner equivalent to that defined by Butcher \& Oemler \shortcite{but78},
is $0.22 \pm 0.05$ \cite{dok00}, comparable to the mean value determined
for clusters at redshifts $0.3 < z < 0.5$.

MS1054$-$03 is clearly an interesting rich cluster with a wealth of
observations over a wide variety of wavelengths and, crucially, with
spectroscopic redshifts for the majority of the objects towards the
cluster centre: currently, spectroscopic redshifts have been measured for
80\% of all galaxies with $19.0 < {\rm I} <22.0$ (the brightest cluster
galaxy has ${\rm I} \sim 19.5$) within the bounds of the HST mosaic. As
such, MS1054$-$03 is an ideal target for a deep cluster radio survey.
Indeed, it was one of the high redshift clusters in the 1.4\,GHz survey of
Stocke \etal\ \shortcite{sto99b}, who detected three radio sources within
the cluster down to a flux density limit of 0.2mJy. These are discussed
later, together with the weaker sources identified in the current
observations.

\section{Observations and data reduction}
\label{vlaobs}

\subsection{Observations and calibration}

MS1054$-$03 was observed using the Very Large Array (VLA) at 5\,GHz in C
array configuration during three 10-hour runs on April 3, 6 and 7 2000. The
total on-source integration time was 96270 seconds. The observations were
carried out simultaneously at two frequencies, 4835 and 4885\,MHz, each
with two circular polarisations and 50\,MHz bandwidth. With this set--up,
the full--width--half--power of the antenna primary beam is about 9
arcmins and the angular resolution about 4 arcsec.  

The observations were carried out using standard VLA procedures. Short
observations of the primary flux calibrator 3C286 (1331+305) were used to
calibrate the flux density scale, assuming flux densities of 7.46 and 7.51
Jy at the two observing frequencies; these are the most recently
determined VLA values, and are approximately 1\% above the flux density
scale of Baars \etal\ \shortcite{baa77}. Observations of 3C286 separated
in time by about 6 hours were used to determine the absolute polarisation
position angle and to estimate the uncertainty in this calibration ($\pm
2^{\circ}$) from the difference between the solutions for the two
different scans. The secondary calibrator 1058+015, offset 5 degrees from
MS1054$-$03, was observed at 30 minute intervals throughout the runs to
provide accurate phase calibration. The wide range of parallactic angles
at which this calibrator was observed enabled the on-axis antenna
polarisation response terms to be accurately determined.

\subsection{Editing and Imaging}

After first discarding data from any antenna or baseline showing excessive
noise (very little for the first run, about 5\% of the data from the
second run and 10\% of the data from the third run), the data were \clean
ed using the \aips\ task \imagr. Then, the presence of a just sufficiently
bright (7mJy) point source towards the centre of the field, enabled two
cycles of phase self--calibration to be carried out, which reduced the map
rms by 10--15\%. Final maps of the field were then produced in the Stokes
parameters I, Q and U, by further \clean ing the datasets. The maps were
produced using an intermediate data weighting between those of natural and
uniform weighting, by setting the data weighting \robust\ parameter to
0.5. This weighting provided the best compromise between the higher
sensitivity of a naturally weighted map and the higher angular resolution
of a uniformly weighted map: the rms noise level was only 8\% higher than
that of natural weighting, and a Gaussian restoring beam of
full--width--half--maximum (FWHM) 5 arcsec (north-south) by 4.5 arcsec
(east--west) was used. A second total intensity map was also made at full
angular resolution (4.0 by 3.9 arcsec FWHM) using uniform weighting to
allow more detailed investigation of the source structural parameters; the
rms noise level was about 70\% higher than than that of the \robust $=
0.5$ map.

\begin{figure*}
\centerline{
\psfig{file=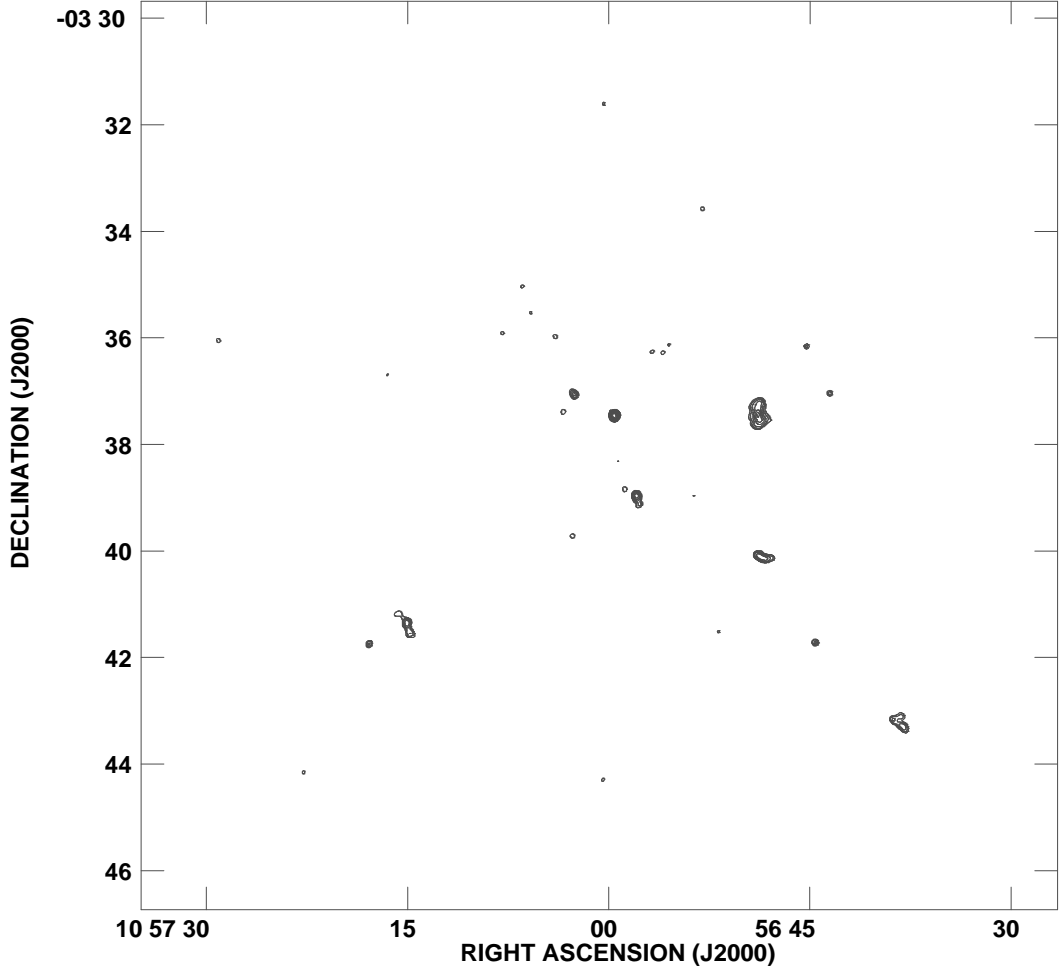,angle=-90,width=14.25cm,clip=}
}
\caption{\label{radiomap} The contour map of the 1024 by 1024 arcsec
field around the centre of the cluster MS1054$-$03. The image rms noise
level is 5.3$\mu$Jy\,beam$^{-1}$, and contours are plotted at ($-$2, $-$1,
1, 2, 4, 8, 16, 32, 64, 128, 256, 512) $\times 32\mu$Jy\,beam$^{-1}$
(6$\sigma$), though note that due to primary beam attenuation (for which
this map has not been corrected) the true sky flux densities are somewhat
higher at large angular separations from the pointing centre: the antennae
primary beam FWHP is about 9 arcmins.}
\end{figure*}

Maps of the field of size 1024 $\times$ 1024 pixels, with pixel size 1.0
arcsec were made. Extra \clean\ boxes were placed around a further six
sources outside of this field, to prevent any increase in the map noise
due to their sidelobes, but these extra sources are not considered further
in this paper since they are well beyond the primary beam for these
observations and are unlikely to be associated with the cluster. 25000
\clean\ iterations were used, which \clean ed the maps down to a level of
$2.0 \sigma$. Although negative \clean\ components were found after a much
smaller number of iterations, this deep \clean ing is essential to fully
remove the source sidelobes. A contour plot of the Stokes I 5 by 4.5
arcsec resolution map of the field is displayed in Figure~\ref{radiomap}.

\subsection{Source extraction and properties}

The resultant rms noise of the lower resolution maps is $5.3 \pm 0.1
\mu$Jy, within a few percent of the theoretically expected value,
indicating the success of the editing and calibration procedures. Real
sources were taken to be those which had either a peak image flux
density\footnote{Note that due to primary beam attenuation the true sky
flux densities are higher than the observed image flux densities at large
radii; see discussion of corrections later in this section. For clarity,
in this paper the term ``sky flux density'' is adopted for the true flux
density of the source and the term ``image flux density'' refers to the
(uncorrected) lower value measured on the images due to primary beam
attenuation.}  in excess of 32\,$\mu$Jy (6$\sigma$), or a peak flux
density in excess of 21\,$\mu$Jy (4$\sigma$) and extended structure whose
integrated image flux density was detected above the 6$\sigma$
level. There are no negative regions which would be selected according to
these criteria if the sign of the radio flux density were inverted across
the entire map, and so it is reasonable to expect that all of the
extracted sources are real. Only those sources within 400 arcsec of the
pointing centre (corresponding to 80\% primary beam attenuation) were
considered. 34 sources were thus selected from the image, and are detailed
in Table~\ref{radprops}.

\begin{table*}
\caption{\label{radprops} The radio properties of the radio sources in the
field. Provided are: (1) the source catalogue name, ordered in increasing
distance from the centre of the cluster; (2,3) the right ascension and
declination of the source; (4) the offset in arcsec of the radio source
from the centre of the cluster, as defined by the position of the
optically brightest cluster galaxy which is at 10 57 00.02, $-$3 37 36.0
(J2000); (5,6) the peak and integrated image flux densities of the radio
source (see text); (7) the sky flux density of the source after correction
for primary beam attenuation; (8) where the redshift of the source is
known (see Table~\ref{radopttab}), the rest--frame 5\,GHz radio luminosity
of the source, assuming a spectral index of 0.8; (9) the nature of the
source (U -- unresolved; R/U -- possible resolved; R -- resolved; E --
extended); (10) the angular size of the source; (11) the position angle of
the source.}
\begin{center}
\begin{tabular}{lrrrrrrccrr}
~~~Source&\multicolumn{2}{c}{RA~~~~(J2000)~~~~Dec}&$\Delta r$&$S_{\rm p}$&
$S_{\rm int}$&$S_{\rm corr}$&log($L_{\rm 5GHz}$)&Nature&$\Theta$&PA\\
~~~~~(1) & (2)~~~~~ & (3)~~~~ & (4) & (5)~~ & (6)~~ & (7)~~ & 
(8)~~~ & (9)~~ & (10) & (11) \\
             &             &              & [$''$]&  [$\mu$Jy] &  [$\mu$Jy]&[$\mu$Jy]&[W\,Hz$^{-1}$]&& [$''$]   &[deg]\\ 
\\
MS1054$-$C1  & 10 56 59.59 & $-$3 37 27.7 &  10.6 &    7302.0. &    7309.0  &    7302.0  & 25.1 & U &   $< 0.5$ & ... \\
             & $\pm 0.00$  & $\pm 0.0$    &       & $\pm  5.3$ & $\pm 10.7$ & $\pm  5.3$ &      &   &           &     \\
MS1054$-$C2  & 10 57 02.03 & $-$3 37 31.4 &  29.9 &      21.3  &      33.2  &      32.1  & 22.7 &R/U&      4.8  &  54 \\
             & $\pm 0.05$  & $\pm 0.7$    &       & $\pm  5.3$ & $\pm  8.9$ & $\pm  8.9$ &      &   &   $< 7.9$ &     \\
MS1054$-$C3  & 10 57 02.60 & $-$3 37 03.6 &  50.5 &     341.0  &     461.0  &     465.7  & 23.9 & R &      3.5  &  36 \\
             & $\pm 0.01$  & $\pm 0.1$    &       & $\pm  5.3$ & $\pm 12.6$ & $\pm 12.7$ &      &   & $\pm 0.2$ &     \\
MS1054$-$C4  & 10 57 03.41 & $-$3 37 23.6 &  53.0 &      51.9  &     228.0  &     230.5  & ...  & E&$\approx 15$~~&140\\ 
             & $\pm 0.06$  & $\pm 0.7$    &       & $\pm  5.3$ & $\pm 24.0$ & $\pm 24.1$ &      & [Diffuse]           \\
MS1054$-$C5  & 10 57 03.80 & $-$3 37 42.7 &  57.0 &      24.2  &      38.2  &      38.8  & 22.8 &R/U&      4.3  &  75 \\
             & $\pm 0.05$  & $\pm 0.4$    &       & $\pm  5.3$ & $\pm  8.4$ & $\pm  8.5$ &      &   &   $< 6.9$ &     \\
MS1054$-$C6  & 10 57 04.09 & $-$3 37 07.8 &  67.0 &      35.3  &      51.5  &      52.8  & 23.0 &R/U&      3.2  & 108 \\
             & $\pm 0.03$  & $\pm 0.4$    &       & $\pm  5.3$ & $\pm  7.1$ & $\pm  7.3$ &      &   &   $< 5.1$ &     \\
MS1054$-$C7  & 10 56 58.86 & $-$3 38 51.4 &  77.4 &      61.1  &      91.6  &      95.0  & ...  &R/U&      1.5  &  41 \\
             & $\pm 0.02$  & $\pm 0.2$    &       & $\pm  5.3$ & $\pm  7.2$ & $\pm  7.5$ &      &   &   $< 2.3$ &     \\
MS1054$-$C8  & 10 56 57.93 & $-$3 38 58.0 &  87.9 &     839.0  &    1358.0  &    1428.0  & 24.4 & E &$\approx 8$~~&148\\
             & $\pm 0.01$  & $\pm 0.1$    &       & $\pm  5.3$ & $\pm 22.6$ & $\pm 24.1$ &      & [Core-jet]          \\
MS1054$-$C9  & 10 56 56.86 & $-$3 36 16.0 &  92.9 &      60.9  &      61.1  &      64.5  & ...  & U &   $< 3.2$ & ... \\
             & $\pm 0.02$  & $\pm 0.2$    &       & $\pm  5.3$ & $\pm  6.4$ & $\pm  6.3$ &      &   &           &     \\
MS1054$-$C10 & 10 56 55.97 & $-$3 36 16.7 &  99.8 &      52.3  &      47.6  &      56.1  & ...  & U &   $< 2.8$ & ... \\
             & $\pm 0.01$  & $\pm 0.2$    &       & $\pm  5.3$ & $\pm  8.0$ & $\pm  6.3$ &      &   &           &     \\
MS1054$-$C11 & 10 56 55.51 & $-$3 36 08.4 & 110.6 &      40.5  &      39.8  &      44.3  & ...  & U &   $< 2.7$ & ... \\
             & $\pm 0.02$  & $\pm 0.3$    &       & $\pm  5.3$ & $\pm  7.3$ & $\pm  6.4$ &      &   &           &     \\
MS1054$-$C12 & 10 57 03.99 & $-$3 35 58.9 & 114.5 &      63.4  &      63.7  &      69.7  & 23.1 & U &   $< 2.5$ & ... \\
             & $\pm 0.01$  & $\pm 0.2$    &       & $\pm  5.3$ & $\pm  6.8$ & $\pm  6.5$ &      &   &           &     \\
MS1054$-$C13 & 10 56 53.70 & $-$3 38 58.0 & 125.2 &      40.6  &      39.4  &      45.8  & ...  & U &   $< 3.5$ & ... \\
             & $\pm 0.02$  & $\pm 0.3$    &       & $\pm  5.3$ & $\pm  6.7$ & $\pm  6.7$ &      &   &           &     \\
MS1054$-$C14 & 10 57 02.73 & $-$3 39 43.2 & 133.5 &      56.0  &      62.2  &      71.7  & 23.1 & U &   $< 4.3$ & ... \\
             & $\pm 0.02$  & $\pm 0.3$    &       & $\pm  5.3$ & $\pm  8.3$ & $\pm  9.6$ &      &   &           &     \\
MS1054$-$C15 & 10 57 00.01 & $-$3 35 13.5 & 142.5 &      28.5  &      72.3  &      85.3  & 22.1 & R &      7.2  & 168 \\
             & $\pm 0.03$  & $\pm 0.5$    &       & $\pm  5.3$ & $\pm 14.2$ & $\pm 16.7$ &      &   & $\pm 1.5$ &     \\
MS1054$-$C16 & 10 57 05.85 & $-$3 35 32.4 & 151.3 &      38.9  &      37.9  &      47.1  & ...  & U &   $< 3.1$ & ... \\
             & $\pm 0.02$  & $\pm 0.3$    &       & $\pm  5.3$ & $\pm  7.2$ & $\pm  7.1$ &      &   &           &     \\
MS1054$-$C17 & 10 57 07.95 & $-$3 35 54.8 & 156.0 &      47.4  &      45.3  &      58.2  & ...  & U &   $< 3.2$ & ... \\
             & $\pm 0.02$  & $\pm 0.3$    &       & $\pm  5.3$ & $\pm  7.0$ & $\pm  7.2$ &      &   &           &     \\
MS1054$-$C18 & 10 56 48.90 & $-$3 37 26.9 & 166.7 &    1029.0  &    6023.0  &    7653.0  & 23.8 & E &$\approx 22$~~& 7\\
             & $\pm 0.01$  & $\pm 0.1$    &       & $\pm  5.3$ & $\pm 36.0$ & $\pm 93.5$ &      & [FR\,I]             \\
MS1054$-$C19 & 10 57 06.46 & $-$3 35 02.2 & 181.5 &      49.7  &      63.7  &      85.3  & ...  &R/U&     2.7   & 125 \\ 
             & $\pm 0.02$  & $\pm 0.2$    &       & $\pm  5.3$ & $\pm  8.5$ & $\pm 11.5$ &      &   &   $< 4.0$ &     \\ 
MS1054$-$C20 & 10 56 48.59 & $-$3 40 07.2 & 228.3 &     485.3  &    1493.0  &    2396.5  & ...  & E &$\approx 11$~~&69\\
             & $\pm 0.01$  & $\pm 0.1$    &       & $\pm  5.3$ & $\pm 23.6$ & $\pm 58.9$ &      & [FR\,I?]            \\
MS1054$-$C21 & 10 56 45.25 & $-$3 36 09.8 & 237.3 &      69.2  &      77.6  &     130.4  & 22.6 & U &   $< 3.8$ & ... \\
             & $\pm 0.02$  & $\pm 0.2$    &       & $\pm  5.3$ & $\pm  8.3$ & $\pm 14.0$ &      &   &           &     \\
MS1054$-$C22 & 10 56 43.53 & $-$3 37 02.6 & 249.0 &     106.0  &     104.5  &     188.3  & ...  & U &   $< 1.6$ & ... \\
             & $\pm 0.01$  & $\pm 0.1$    &       & $\pm  5.3$ & $\pm  6.1$ & $\pm 10.5$ &      &   &           &     \\
MS1054$-$C23 & 10 57 16.50 & $-$3 36 41.7 & 252.6 &      39.4  &      49.9  &      90.1  & ...  & U &   $< 4.3$ & ... \\ 
             & $\pm 0.03$  & $\pm 0.4$    &       & $\pm  5.3$ & $\pm  7.5$ & $\pm 13.7$ &      &   &           &     \\
MS1054$-$C24 & 10 57 16.53 & $-$3 38 45.1 & 256.6 &      33.8  &      38.1  &      70.2  & ...  & U &   $< 4.6$ & ... \\
             & $\pm 0.03$  & $\pm 0.4$    &       & $\pm  5.3$ & $\pm  7.6$ & $\pm 14.0$ &      &   &           &     \\
MS1054$-$C25 & 10 56 53.02 & $-$3 33 34.6 & 263.2 &      56.6  &      58.7  &     107.4  & ...  & U &   $< 3.6$ & ... \\
             & $\pm 0.02$  & $\pm 0.3$    &       & $\pm  5.3$ & $\pm  5.9$ & $\pm 10.1$ &      &   &           &     \\
MS1054$-$C26 & 10 56 51.88 & $-$3 41 31.4 & 265.1 &      39.9  &       41.8 &      76.4  & ...  & U &   $< 3.8$ & ... \\
             & $\pm 0.02$  & $\pm 0.4$    &       & $\pm  5.3$ & $\pm  6.2$ & $\pm 10.1$ &      &   &           &     \\
MS1054$-$C27 & 10 56 50.23 & $-$3 33 42.9 & 275.4 &      29.6  &      71.4  &     143.9  & ...  & R &      6.3  & 70  \\
             & $\pm 0.04$  & $\pm 0.6$    &       & $\pm  5.3$ & $\pm  9.7$ & $\pm 19.6$ &      &   & $\pm 2.4$ &     \\
\\
\end{tabular}	
\end{center}
\centerline {\it continued on following page}
\end{table*}

\addtocounter{table}{-1}
\begin{table*}
\caption{\it continued from previous page}
\begin{tabular}{lrrrrrrccrr}
Source&\multicolumn{2}{c}{RA~~~~(J2000)~~~~Dec}&$\Delta r$&$S_{\rm p}$&$S_{\rm int}$&$S_{\rm corr}$ & 
log($L_{\rm 5GHz}$)&Nature&$\Theta$&PA\\
             &             &              & [$''$]&  [$\mu$Jy] &  [$\mu$Jy]&[$\mu$Jy]&[W\,Hz$^{-1}$]&&[$''$]    &[deg]\\ 
\\
MS1054$-$C28 & 10 56 46.09 & $-$3 41 32.7 & 315.4 &      32.0  &      30.8  &      79.9  & ...  & U &   $< 3.2$ & ... \\
             & $\pm 0.03$  & $\pm 0.4$    &       & $\pm  5.3$ & $\pm  5.6$ & $\pm 13.3$ &      &   &           &     \\
MS1054$-$C29 & 10 56 39.37 & $-$3 38 49.3 & 317.7 &      30.0  &      41.9  &     106.3  & ...  &R/U&      4.1  & 23  \\
             & $\pm 0.03$  & $\pm 0.6$    &       & $\pm  5.3$ & $\pm  7.9$ & $\pm 20.0$ &      &   &   $< 5.7$ &     \\
MS1054$-$C30 & 10 57 15.07 & $-$3 41 21.2 & 318.6 &     550.7  &    1230.8  &    3131.8  & ...  & E &$\approx 28$~~&11\\
             & $\pm 0.01$  & $\pm 0.1$    &       & $\pm  5.3$ & $\pm 32.1$ &$\pm 124.8$ &      & [FR\,I (II?)]       \\
MS1054$-$C31 & 10 56 44.61 & $-$3 41 43.3 & 338.2 &     167.6  &     168.3  &     484.4  & ...  & U &   $< 2.5$ & ... \\
             & $\pm 0.01$  & $\pm 0.1$    &       & $\pm  5.3$ & $\pm  6.1$ & $\pm 23.4$ &      &   &           &     \\
MS1054$-$C32 & 10 57 00.36 & $-$3 31 36.0 & 360.1 &      45.2  &      47.3  &     153.2  & ...  & U &   $< 3.2$ & ... \\
             & $\pm 0.02$  & $\pm 0.3$    &       & $\pm  5.3$ & $\pm  6.2$ & $\pm 18.0$ &      &   &           &     \\
MS1054$-$C33 & 10 57 17.88 & $-$3 41 44.9 & 365.3 &     166.2  &     174.0  &     589.4  & ...  & U &   $< 3.0$ & ... \\
             & $\pm 0.01$  & $\pm 0.1$    &       & $\pm  5.3$ & $\pm  9.4$ & $\pm 30.3$ &      &   &           &     \\
MS1054$-$C34 & 10 57 23.00 & $-$3 35 15.0 & 371.8 &      26.8  &     424.0  &    1600.0  & ...  & E&$\approx 60$~~&110\\
             & $\pm 0.60$  & $\pm 5.0$    &       & $\pm  5.3$ & $\pm 56.0$ &$\pm 200.0$ &      & [Diffuse]           \\
\end{tabular}
\end{table*}

The absolute position of each source was determined, together with its
position relative to the cluster centre (as determined by the optically
selected brightest cluster galaxy). For extended sources, note that the
quoted position is the position of the peak of the radio flux, and may not
accurately correspond to the location of the associated optical
galaxy. Similarly for the very faintest sources the theoretically
determined errors may be somewhat optimistic.  

The peak flux density of each source was measured directly from the image,
and corrected for the effects of chromatic aberration: the 50\,MHz
bandwidth leads to radial smearing of a true point source, and results in
the measured peak flux density being decreased by 5\% at 230 arcsec
radius, and 10\% at 345 arcsec radius for these observations. For
unresolved or barely resolved sources, the integrated flux density was
calculated from a Gaussian fit using the \aips\ task \jmfit. Sources which
are clearly extended, or of such low signal--to--noise that Gaussian
fitting may be misleading, had their integrated flux density calculated by
summing the flux density within a small box placed around the source. No
correction has been made for surface brightness biases (e.g. Pascarelle
et~al. 1998)\nocite{pas98}, which should be fairly negligible at the
relatively low angular resolution of these observations.  The image flux
densities were then corrected for primary beam attenuation to determine
the true sky flux density of each source. This attenuation is
approximately 50\% at a radius of 270 arcsec. For those sources which were
either extended or resolved, the integrated flux density was taken and
corrected to determine the true sky flux density; for the unresolved
sources, the peak flux density was used. All three flux densities are
provided for each source in Table~\ref{radprops}.

Morphological parameters of each source were determined using \jmfit: for
the sources brighter than $60 \mu$Jy the higher angular resolution
uniformly weighted map was used for this, with the lower resolution higher
sensitivity map being used for the weaker sources. The deconvolved angular
size of each source was determined, or an upper limit set thereon, after
correction for radial smearing due to the non-zero bandwidth. Each source
was classified as either: (i) unresolved (U), if only an upper limit could
be set to its angular size; (ii) possibly resolved (R/U) if an angular
size was determined but the one sigma lower limit was below zero; (iii)
resolved (R), if the source was clearly resolved; or (iv) extended (E), if
the intensity distribution could not be fit using a single Gaussian
profile. In the later case, some indication of the nature of the radio
source is also provided in the table. It should be noted here that no
attempt has been made to correct for any overestimation of the angular
sizes that may occur due to the presence of noise when fitting small radio
sources detected at only the few sigma level (e.g. Windhorst \etal\
1984)\nocite{win84}. This effect is negligible for the brighter sources,
but could be $\gta 10\%$ for the weakest sources: as a guide, the angular
sizes of the fainter sources in the sample are typically about 15\% higher
if determined from the 5 arcsec resolution image than from the 4 arcsec
resolution image used.

\section{The {$\mathbf \mu$}Jy radio source population}
\label{radpop}

\subsection{Comparison with background counts}

Numerous field surveys have been carried out of the $\mu$Jy radio source
population to determine the differential radio source counts at different
radio frequencies (see introduction). Fomalont \etal\
\shortcite{fom91} have derived a parameterisation for the integral source
count at 5\,GHz of: $N(>S) = (23.2 \pm 2.8) S^{-1.18 \pm 0.19}$, where
$N(>S)$ is the number of sources per square arcmin with a total flux
density greater than $S$, measured in $\mu$Jy. This formula can be used to
predict the mean number of background sources that is expected in a 400
arcsec radius circle reaching a flux density limit of $32\mu$Jy at the
centre, taking into account the variation in the sensitivity with radius
due to primary beam attenuation. That gives a background prediction of
about 25 ($\pm 5$) sources which, compared to the 34 sources observed,
suggests that of order a quarter of the detected sources are likely to be
associated with the cluster.

\begin{figure}
\centerline{
\psfig{file=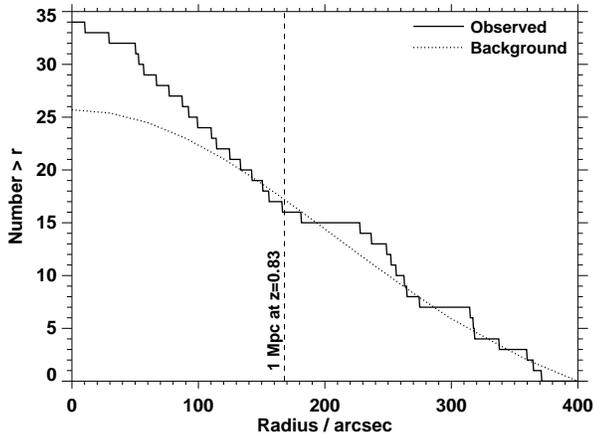,angle=90,width=8.4cm,clip=}
}
\caption{\label{raddensplot} The cumulative distributions (working from a
radius of 400 arcsec in to the centre of the cluster) of detected radio
sources (solid line) compared to the predicted background counts (dotted
line). The observed distribution is fully compatible with expectations at
large radii, but a clear excess of sources, significant at the $>99$\%
significance level, is detected at small radii. These excess counts
correspond to the radio sources associated with the cluster, and are found
within a radial distance of order 700\,kpc (evaluated at $z=0.83$).}
\end{figure}

This analysis can be taken a step further by investigating the radial
distribution of radio source counts in the cluster. In
Figure~\ref{raddensplot} the number of detected sources is compared to the
expected number of background sources as a function of distance from the
centre of the cluster. A clear ($>99$\% significant) excess of sources is
detected in the inner 2 arcmin radius, corresponding to about 700\,kpc at
the redshift of the cluster. Beyond this radius, the number counts of
detected sources are fully consistent with background predictions. These
results are also provided numerically in a number of radial annuli in
Table~\ref{raddens}.

\begin{table}
\caption{\label{raddens} The distribution of radio sources with distance
from the cluster centre, compared to the expected background counts. For
each radius range (1) the table provides the mean radius (2), the limiting
sky flux density at that radius (3; this corresponds to a limiting image
flux density of 32$\mu$Jy corrected for the primary beam attenuation
factor), the mean number of background sources per square arcmin expected
above this flux density (4), the total number of background sources
expected to be detected in this radius range, together with the expected
variation assuming Poisson statistics (5), and the true number of sources
detected (6). A clear excess of detected sources is seen in the inner 2
arcmins, which corresponds to about 700\,kpc at the redshift of the
cluster (see also Figure~3).}
\begin{tabular}{cccccc}
Radius & $\overline{R}$ & $\overline{S_{\rm lim}}$ & $N(>S_{\rm lim})$ &
$N_{\rm bk}$ & $N_{\rm obs}$ \\
~[$''$]   & [$''$] & [$\mu$Jy] &[arcmin$^{-2}$]&                 \\
    (1)    &    (2)   &    (3)    &     (4)       &     (5)        & (6) \\
  0 -- 60  &     42   &    32.0   &    0.389      &  $1.2 \pm 1.1$ & 5  \\
 60 -- 120 &     95   &    33.9   &    0.363      &  $3.4 \pm 1.8$ & 7  \\
120 -- 180 &    153   &    38.7   &    0.310      &  $4.9 \pm 2.2$ & 6  \\
180 -- 240 &    212   &    48.0   &    0.241      &  $5.3 \pm 2.3$ & 3  \\
240 -- 300 &    272   &    63.0   &    0.175      &  $4.9 \pm 2.2$ & 6  \\
300 -- 360 &    331   &    87.7   &    0.118      &  $4.1 \pm 2.0$ & 4  \\
360 -- 400 &    381   &   129.1   &    0.075      &  $2.0 \pm 1.4$ & 3  \\
\end{tabular}
\end{table}

\subsection{Radio--optical comparisons: the optical data}
\label{radopt}

As discussed in Section~\ref{ms1054}, an HST WFPC2 mosaic of the
MS1054$-$03 cluster has been made covering an area of 5 by 5 arcminutes
for an exposure time of 6500 second per pointing per filter in each of the
F606W ($R$) and F814W ($I$) filters. This reaches a depth in the $I$ band
of $I \sim 26$ for a point source.  VLT observations using FORS in the
$U$, $B$ and $V$ wavebands have also been taken, covering a sky area of
6.8 by 6.8 arcmins and reaching a point source limiting depth (3$\sigma$)
in the $V$ band of $V \sim 27$ (Franx \etal\ in preparation).

To compare the radio and optical data, the two datasets must be placed in
the same astrometric frames. Radio observations are automatically in the
International Celestial Reference Frame (ICRF), with positional errors
significantly below 0.1 arcsec, but the absolute positions of optical
images can be uncertain at the arcsec level. Initial astrometry of the
optical frames was taken from the HST astrometric headers, after
correction for geometric distortions \cite{dok00}. This astrometry was
then improved by considering the relative radio and optical positions of
those unresolved radio sources which had unambiguous optical counterparts
and adjusting the optical astrometry to produce the minimum combined
radio--optical offsets for these sources, taking into account the
uncertainties on the optical and radio positions of each source. This
improvement resulted in a small offset of the optical frames by 0.17
arcsec E and 0.13 arcsec N from the astrometry of van Dokkum \etal, with
the central cluster galaxy now having a position of 10 57 00.02, $-$3 37
36.0 (J2000). The resultant optical frame astrometry should be within 0.2
arcsec of the ICRF. The magnitudes and positions of the optical galaxies
were obtained using SExtractor \cite{ber96} on the HST image mosaic
\cite{dok00}.

\begin{table*}
\caption{\label{radopttab} Comparisons between the radio sources and the
optical data. For each source, the offset (in arcsec) of the radio source
from the central cluster galaxy (at 10 57 00.02, $-$3 37 36.0, J2000) in
RA (positive is West) and Dec (positive is North) is given. For those
sources which lie within the fields of the optical imaging, details of any
optical galaxy brighter than I$=26$ within 2.5 arcsec of the radio
position are also given. The old identification numbers of each galaxy are
taken from the original numbering scheme for the HST imaging observations
(van Dokkum \etal\ 2000), with a value of $-$99 indicating a galaxy
previously without a number. The new identification numbers represent a
reclassification including all galaxies on the HST frame, introduced for
the spectroscopic observations: these new identification numbers will be
used in all forthcoming papers. The offsets $\delta x$ and $\delta y$ are
from the central cluster galaxy, as defined for the radio sources, and the
`Offset' column gives the positional difference (in arcsec) between the
radio and optical locations. The morphological type of the galaxy is given
where known, as is its magnitude through the F814W ($\sim I$) filter of
the HST and the F606W$-$F814W ($\sim R-I$) colour for those objects within
the HST mosaic. The penultimate column gives the source redshift, if
measured, and the final column indicates with a `C' the confirmed cluster
galaxies, and with `P' the potential galaxy pairs.}
\begin{tabular}{lrrrrrrcccccc}
Source & \multicolumn{2}{c}{Radio Source}& 
\multicolumn{10}{c}{..................Optical Galaxies within 2.5 arcsec..................} \\  
&$\delta x$~~&$\delta y$~~&Old Id.&New Id.&$\delta x$&$\delta y$& 
Offset&Type& F814W  &F606W$-$F814W&  $z$ & Note  \\  
\\
MS1054-C1  &     6.4 &     8.3 & 1459 & 4705 &     6.1 &     8.3 & 0.3 & Sc & 20.66 &  1.71 & 0.8452 & C \\
MS1054-C2  & $-$30.1 &     4.6 & 1435 & 5442 & $-$29.7 &     4.5 & 0.4 &S0/a& 21.87 &  1.96 & 0.8195 & CP\\
           &         &         & 1431 & 5432 & $-$29.4 &     2.3 & 2.4 &Sa/b& 22.11 &  1.59 & 0.8148 & CP\\
           &         &         & 1433 & 5465 & $-$32.3 &     5.7 & 2.5 &--- & 22.63 &  1.91 & 1.0760 & \\
MS1054-C3  & $-$38.6 &    32.4 & 1656 & 5529 & $-$37.8 &    31.2 & 1.4 &E/S0& 21.06 &  2.09 & 0.8216 & CP\\
           &         &         & 1655 & 5534 & $-$37.9 &    34.1 & 1.9 & E  & 20.94 &  1.86 & 0.8397 & CP\\
MS1054-C4  & $-$50.7 &    12.4 & ---  &  --- &   ---   &     --- & --- & ---&  ---  &   --- &   ---  \\
MS1054-C5  & $-$56.6 &  $-$6.7 & 1403 & 5599 & $-$56.8 &  $-$5.5 & 1.2 & Sc & 20.14 &  1.31 & 0.8113 & C\\
MS1054-C6  & $-$61.0 &    28.2 & 1584 & 6038 & $-$61.5 &    28.3 & 0.5 & E  & 20.76 &  2.13 & 0.8311 & CP\\
           &         &         &$-$99 & 6143 & $-$59.4 &    27.2 & 1.9 & ---& 23.49 &  0.55 &   ---  \\
           &         &         &$-$99 & 6164 & $-$63.2 &    28.3 & 2.2 & E  & 21.49 &  2.11 &   ---  & P \\ 
           &         &         &$-$99 & 6145 & $-$59.6 &    30.2 & 2.5 & ---& 22.80 &  0.65 &   ---  \\ 
MS1054-C7  &    17.4 & $-$75.4 &  735 & 4362 &    17.4 & $-$74.9 & 0.5 &--- & 21.82 &  0.88 &   ---  \\
MS1054-C8  &    31.2 & $-$82.0 &$-$99 & 4012 &    31.0 & $-$82.8 & 0.8 &S0/a& 21.50 &  1.98 & 0.8297 & CP\\
           &         &         &  710 & 3995 &    30.5 & $-$81.0 & 1.2 & E  & 20.78 &  2.14 & 0.8342 & CP\\
MS1054-C9  &    47.3 &    80.0 & 1865 & 3456 &    49.5 &    79.0 & 2.4 &Star& 17.82 &  1.85 &  ---  \\
MS1054-C10 &    60.6 &    79.3 & 1851 & 3104 &    60.8 &    78.9 & 0.4 &--- & 23.53 &  0.78 &  ---   \\
           &         &         & 1851 & 3070 &    60.0 &    79.6 & 0.6 &--- & 23.44 &  1.07 &  ---   \\
MS1054-C11 &    67.5 &    87.6 & 1895 & 2847 &    67.8 &    87.7 & 0.3 &--- & 23.30 &  0.71 &  ---   \\
           &         &         &$-$99 & 2825 &    67.9 &    85.2 & 2.5 &--- & 24.62 &  0.56 &  ---   \\
MS1054-C12 & $-$59.4 &    97.9 & 1942 & 5756 & $-$59.3 &    97.9 & 0.1 & E  & 20.98 &  2.09 & 0.8309 & C \\
MS1054-C13 &    94.6 & $-$82.0 &$-$99 & 2036 &    94.1 & $-$82.2 & 0.5 &Sa/b& 21.92 &  1.63 &  ---   \\
           &         &         &$-$99 & 1943 &    96.3 & $-$80.2 & 2.5 &--- & 23.87 &  0.76 &  ---   \\
MS1054-C14 & $-$40.6 &$-$127.2 &$-$99 &  564 & $-$41.1 &$-$127.5 & 0.6 &--- &  ---  &  ---  & 0.8321 & C \\
MS1054-C15 &     0.1 &   142.5 & 2132 & 4416 &  $-$0.6 &   142.9 & 0.7 & Sc & 17.96 &  1.23 & 0.2499 \\
MS1054-C16 & $-$87.3 &   123.6 & ---  &  --- &   ---   &     --- & --- & ---&  ---  &   --- &   ---  \\
MS1054-C17 &$-$118.7 &   101.2 &$-$99 & 7702 &$-$118.3 &   101.7 & 0.6 & ---& 25.91 &   --- &   ---  \\
MS1054-C18 &   166.5 &     9.1 & 1592 &   1  &   166.6 &     9.7 & 0.6 &  E & 16.54 &  0.85 & 0.1830 \\
MS1054-C19 & $-$96.4 &   153.8 & 2151 &  --- & $-$95.9 &   154.0 & 0.5 & ---&  ---  &   --- & ---    \\
MS1054-C20 &   171.1 &$-$151.2 &  519 &  --- &   171.2 &$-$152.8 & 1.6 & ---&  ---  &   --- & ---    \\
MS1054-C21 &   221.1 &    86.2 &$-$99 &   8  &   221.6 &    85.4 & 0.9 &E/S0& 18.56 &  1.21 & 0.3740 \\
MS1054-C22 &   246.8 &    33.4 &$-$99 &  --- &   246.7 &    32.7 & 0.7 & --- & ---  &   --- & ---    \\
MS1054-C23 &$-$246.7 &    54.3 &
\multicolumn{9}{c}{..........................Outside optical fields..........................}\\
MS1054-C24 &$-$247.1 & $-$69.1 & 
\multicolumn{9}{c}{..........................Outside optical fields..........................}\\
MS1054-C25 &   104.8 &   241.4 & ---  &  --- &   ---   &   ---   & --- & ---&   --- &   --- &  ---   \\
MS1054-C26 &   121.9 &  -235.4 & 
\multicolumn{9}{c}{..........................Outside optical fields..........................}\\
MS1054-C27 &   146.6 &   233.1 &$-$99 &  --- &   147.4 &  232.2  & 1.2 & ---&   --- &   --- &  ---   \\
MS1054-C28 &   208.5 &$-$236.7 &
\multicolumn{9}{c}{..........................Outside optical fields..........................}\\
MS1054-C29 &   309.1 & $-$73.3 &$-$99 &  --- &   308.7 & $-$73.1 & 0.5 & ---&   --- &   --- &  ---   \\
MS1054-C30 &$-$225.3 &$-$225.2 &
\multicolumn{9}{c}{..........................Outside optical fields..........................}\\
MS1054-C31 &   230.7 &$-$247.3 &
\multicolumn{9}{c}{..........................Outside optical fields..........................}\\
MS1054-C32 &  $-$5.1 &   360.0 &
\multicolumn{9}{c}{..........................Outside optical fields..........................}\\
MS1054-C33 &$-$267.4 &$-$248.9 &
\multicolumn{9}{c}{..........................Outside optical fields..........................}\\
MS1054-C34 &$-$344.0 &   141.0 &
\multicolumn{9}{c}{..........................Outside optical fields..........................}\\
\end{tabular}
\end{table*}

\subsection{Radio--optical correlations}
\label{radoptcorrel}

The location of each radio source relative to the centre of the cluster is
given in Table~\ref{radopttab}, together with details of any optical
galaxies with $I<26$, that lie within 2.5 arcsec of the radio source. If
multiple optical candidates are found, only those within 3 magnitudes of
the brightest galaxy are listed, to minimise potential galaxy deblending
problems.  The 2.5 arcsec radius considered is larger than the 0.2 arcsec 
uncertainty in the astrometric frame discussed above, primarily due to 
uncertainties in both the radio and optical positions. 

Contour plots of the 34 radio sources are displayed in
Figure~\ref{overfigs} in the appendix, overlaid where possible upon an
optical image of the field. The nature of each individual source is also
discussed there.  From these results it is seen that of the 34 radio
sources, 25 lie within the fields of the optical imaging. All but four
(MS1054$-$C4, C9, C16, C25) of these sources have optical counterparts
which are very probably associated with the radio source. 10 radio sources
are associated with optical galaxies with magnitudes typical of the bright
cluster galaxies ($19 \lta I \lta 22$), and a further 11 associated with
galaxies whose magnitudes are either unmeasured or fall outside of this
range. It should be noted that although towards the very centre of the
cluster the sky density of detected sources is high, with of order 200
galaxies with $I \lta 26$ detected in the centre-most square arcmin (1
galaxy per 18 square arcsec), in this region all of the radio--optical
coincidences are with bright optical galaxies ($I < 22$). At these
magnitudes the probability of an optical galaxy being at the observed
radio--optical offsets is very low ($\ll 5$\%; from $P(<r) = 1 - {\rm exp}
(-\pi \rho r^2)$, where $\rho$ is the sky density of optical sources);
however, even in the cluster outskirts the probability of chance
coincidence with an optical galaxy of $I\sim 25$ is as high as $\sim
25$\%. It should also not be forgotten that any true radio source hosts
with $I>26$ may lie within 2.5 arcsec of a brighter galaxy, and so be
misidentified.

\section{The cluster radio source population}
\label{hostgals}

Of the 21 sources with optical identifications, spectroscopic redshifts
are available for 11 of the host galaxies, and 8 (assuming here that
MS1054$-$C2 is associated with one or both of the cluster galaxies; see
appendix) of these lie at the redshift of the cluster. This section
concentrates upon these 8 cluster radio sources. Figure~3 provides a
detailed picture of the radio--optical correlations and optical
morphologies of these galaxies.

\begin{figure*}
\centerline{
\psfig{file=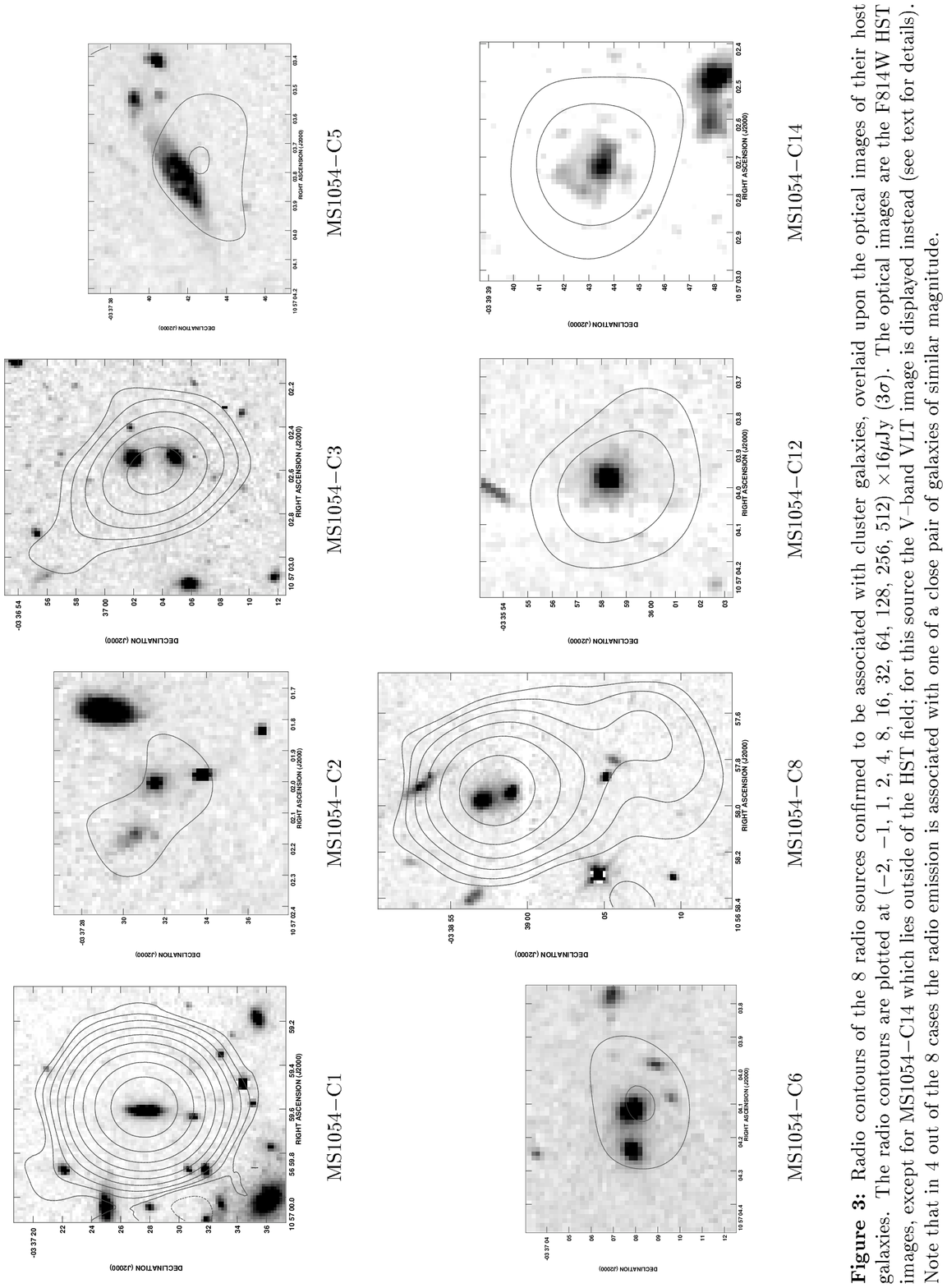,angle=0,width=\textwidth,clip=}
}
\end{figure*}
\addtocounter{figure}{1}

It should be noted that although only 8 of the 34 radio sources are
confirmed to be associated with the cluster, these eight galaxies are
likely to comprise the majority of the cluster radio population brighter
than 32$\mu$Jy: the confirmed cluster galaxies stand out from the
background population not only through the excess radio source counts and
their location in the inner 700\,kpc radius from the cluster centre, but
also by their association with galaxies of magnitudes $20 < I <
22$ and red ($R-I \gg 1$) colours (see Table~\ref{radopttab});
only two of the radio sources without redshifts (MS1054$-$C7 and C13) are
associated with galaxies with measured magnitudes in this range, of which
only C13 is red. Therefore conclusions based upon the properties of these
8 can be interpreted more generally.

A approximate comparison between the radio luminosities and star formation
rates of (non-AGN) cluster galaxies can be made by converting between the
type II supernova rate and the radio emission, since star--forming regions
are transparent at radio wavelengths: $L_{\nu} \approx 1.3 \times 10^{23}
\nu^{-\alpha} n_{\rm SN}$\,W\,Hz$^{-1}$, where $\nu$ is measured in GHz,
$\alpha \simeq 0.8$ is the radio spectral index and $n_{\rm SN}$ is the
supernova rate per year \cite{con90}\footnote{For reference, the Milky Way
has $n_{\rm SN} \simeq 0.041$\,yr$^{-1}$ \cite{tam82} and M82 has $n_{\rm
SN} \simeq 0.1$\,yr$^{-1}$ \cite{con92}.}. Andersen \& Owen
\shortcite{and95} showed that radio sources in nearby clusters are offset
from the radio--far--infrared relationship of field galaxies by typically
a factor of two, and suggest that this is due to a boosting of the radio
emission of cluster galaxies (for a given star formation rate) due to
compression of the magnetic field due to the motion of the galaxy through
the intracluster medium; a more detailed study by Miller \& Owen
\shortcite{mil01} indicates that this applies only for a subsample of
star-forming galaxies, is more prevalent in the richest clusters, and that
the radio enhancement is typically a factor of 2 to 3. Based upon these
results we include an extra factor of two into the above equation to
account for this effect.

For a Salpeter initial mass function with upper and lower mass cut-offs of
125 and 0.15\,$M_{\odot}$ respectively, and assuming that all stars more
massive than 8\,$M_{\odot}$ produce supernovae, there is approximately one
supernova for each 120\,$M_{\odot}$ of star formation. Putting these
together gives a relationship between the radio luminosity and the mean
star formation rate over the past $\sim 10^8$ yrs (the lifetime of the
relativistic synchrotron electrons), $\dot{M}_*$, of:

\begin{equation}
\label{radstar}
\frac{\dot{M}_*}{M_{\odot}{\rm yr}^{-1}} \approx 4.5 \left(\frac{\rm GHz}{\nu}
\right)^{-\alpha} \frac{L_{\nu}}{10^{22} {\rm \,W\,Hz}^{-1}}. 
\end{equation}

In the following subsections this relation will be used to compare the
radio source population with various other classes of cluster galaxies.
In subsection~\ref{butoem} the comparison between the cluster radio
sources and the Butcher--Oemler population is considered. The correlation
of the cluster radio sources with interacting and merging galaxies is
investigated in subsection~\ref{galmerg}, and the correlation with
emission line strengths is discussed in subsection~\ref{emisstren}. The
cluster radio source radio luminosity distribution is discussed in
subsection~\ref{lumdist}.

\subsection{Correlations with galaxy colours}
\label{butoem}

As discussed in the introduction, an increased proportion of bluer
`Butcher Oemler' galaxies, which are believed to have on--going or
recently terminated star formation, are found in high redshift
clusters. To investigate the radio emission from this `Butcher Oemler'
population in MS1054$-$03, the 10 cluster galaxies with colours bluer than
$(U-B)_z = 0.1$ (excluding the blue Sc galaxy MS1054$-$C5 which already
lies within the radio source sample) were selected from the
colour--magnitude diagrams of van Dokkum \etal\ \shortcite{dok00}, and the
mean radio emission from these was calculated.  Only one of the galaxies
was individually detected above the $2\sigma$ level, and that only at
$2.1\sigma$; the combined flux density from all ten galaxies was $2.4 \pm
1.6 \mu$Jy, again only a 1.5$\sigma$ level detection. The measured mean
radio flux density corresponds to $6 \pm 4 M_{\odot}$\,yr$^{-1}$, which is
consistent with the star formation rates necessary to bluen the galaxy
colours by this amount. The current radio observations are not deep enough
to detect typical star--forming galaxies, but only extreme starbursts or
AGN activity.

\begin{figure}
\centerline{
\psfig{file=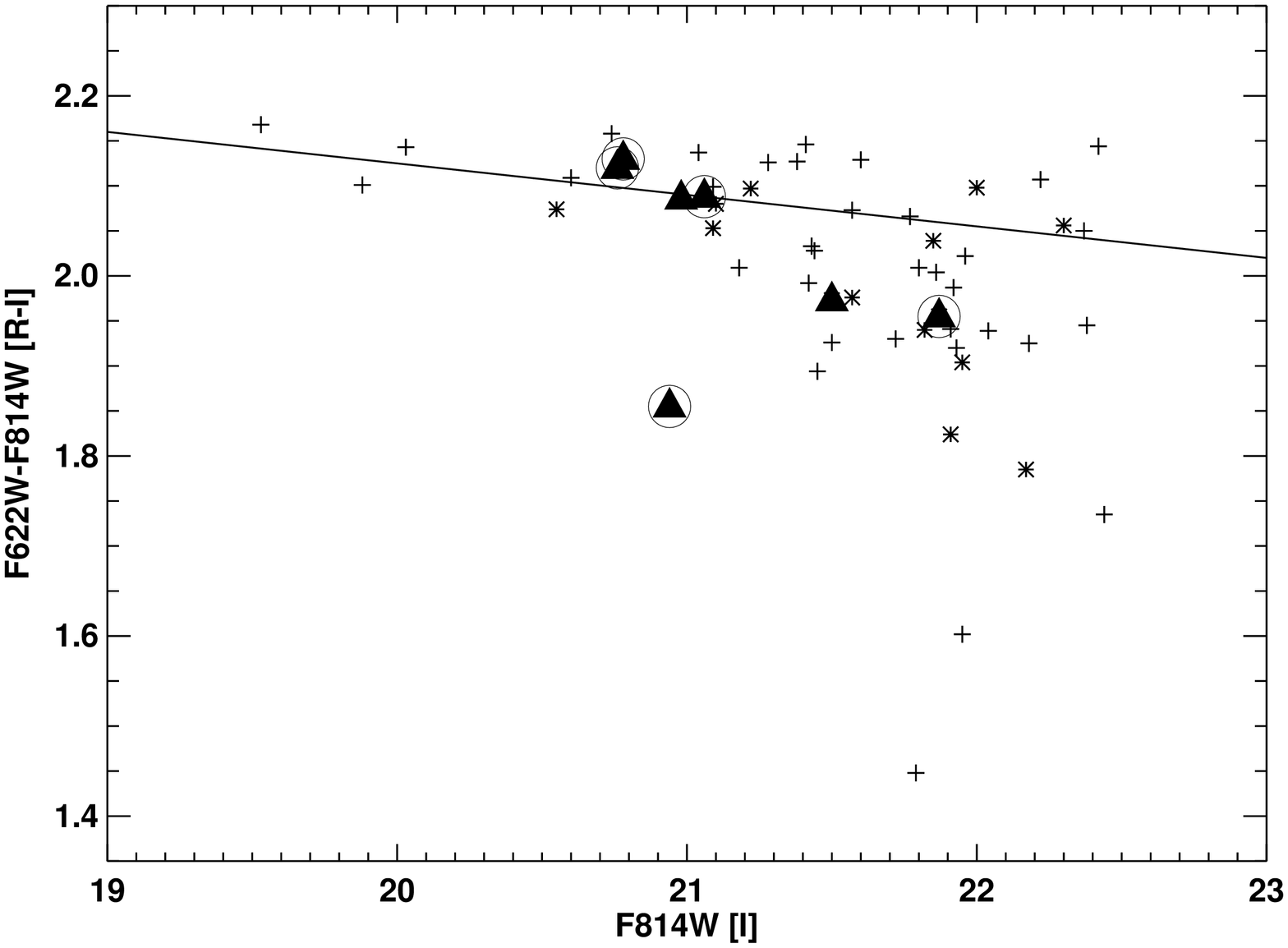,angle=90,width=8cm,clip=}
}
\caption{\label{colmagdiag} The F606W$-$F814W ($\approx R-I$) versus F814W
colour--magnitude relation for the early--type galaxies in the MS1054
cluster. Crosses represent ellipticals and S0s and asterisks are S0/a
galaxies. Those galaxies associated with a radio source are indicated by
filled triangles (in the case of close pairs, both of the pair of galaxies
are indicated). The large circles indicate those radio galaxies which are
one of a close pair. The line represents the fit to the colour--magnitude
relation derived by van Dokkum \etal\ (2000). With only one exception the
galaxies associated with the radio sources have colours consistent with
the tight colour--magnitude relation defined by the other early--type
cluster galaxies.}
\end{figure}

Approaching this issue from the other side, in Figure~\ref{colmagdiag} is
presented an F814W$-$F606W ($\approx R-I$) vs F814W ($I$)
colour--magnitude diagram for all of the early--type confirmed cluster
galaxies, with the radio galaxies indicated as filled triangles, and the
close pairs further indicated by a large circle.  Excepting galaxy 5534
(one of the close pair of galaxies associated with MS1054$-$C3) all of the
galaxies associated with the radio sources have colours consistent with
the colour--magnitude relation of the other early--types in the
cluster. This suggests that their radio emission is not due to starburst
activity, which would tend to bluen the galaxy colours, but rather with
hidden AGN activity.

\subsection{Correlations with galaxy--galaxy interactions}
\label{galmerg}

Of the 8 cluster radio sources, 4 are associated with pairs of cluster
galaxies with projected separations of 10--25\,kpc (1--2.5 arcsec), that
are plausibly undergoing interactions (although in the case of MS1054$-$C3
it must be noted that the velocity difference between the two galaxies in
the pair is over 5000\,km\,s$^{-1}$, and so this must either be a chance
projection or, if an interaction, the galaxy pair is unbound; for
MS1054$-$C6 although the second galaxy is not a spectroscopically
confirmed cluster member, its magnitude and colour are very strongly
indicative that it is). The galaxies comprising these pairs are of type
S0/a or earlier in 7 of the 8 cases, and are within one magnitude of each
other in luminosity. The probability of a chance projection at these
magnitudes and separations is $\ll 5$\%, as discussed above.

Despite the apparent high proportion of `interaction--driven' radio
sources in the cluster, and the fact that the cluster is known to contain
a very high proportion of confirmed binary galaxies and on--going mergers
(galaxy nuclei separated by $\ll 10$\,kpc), not a single one of the radio
sources is associated with any of the 13 cluster objects classified
`merger\,/\,peculiar' by van Dokkum \etal\ \shortcite{dok00}. Indeed, none
of these merging galaxies is detected above the 2$\sigma$ level in the
radio map. This sets interesting limits on the radio luminosity associated
with the cluster merger events: to provide tighter constraints, the radio
emission was averaged across the locations of the 13 confirmed on-going
mergers in the cluster, giving a mean flux density of $2.2 \pm 1.5 \mu$Jy,
only a 1.5$\sigma$ detection. Realistically then, the upper limit to the
radio emission of the merger events is of order 5$\mu$Jy which, at the
redshift of the cluster, corresponds to a rest--frame 5\,GHz radio
luminosity of about $10^{22}$W\,Hz$^{-1}$. From equation 1, this sets a
limit to any starburst accompanying the merger events in the MS1054
cluster, to a mean of order 10$M_{\odot}$\,yr$^{-1}$. This suggests that
the very red colours generally seen for the merging galaxies \cite{dok99a}
are simply the intrinsic colours of the galaxies, rather than being caused
by a massive starburst heavily obscured by dust.

The high proportion of galaxy pairs among radio sources in the cluster and
the absence of radio emission from on-going mergers implies that galaxy
interactions may be more efficient than mergers at inducing radio
activity. It is interesting to compare this result with that of Roche \&
Eales \shortcite{roc00} who found that the majority of a sample of
powerful low redshift 3CR radio galaxies were visibly interacting, but
that the mean projected separation of the interacting pairs was $44 \pm
10$\,kpc. This is about three times larger than that of interacting
ultraluminous infrared galaxies. From this they concluded that radio
bursts occurred at a much earlier stage of an interaction\,/\,merger than
the peak of the starburst activity.

\subsection{Correlations with emission line luminosity}
\label{emisstren}

\subsubsection{Line emission from star formation or AGN activity}

In the absence of high resolution radio observations or radio spectral
indices to distinguish starbursting galaxies from AGN, a separation
between the two classes can still be made based upon to ratio of the radio
to the emission line luminosity of the host galaxies. When combined with
optical morphologies and optical spectral properties this division can be
made with a fair degree of accuracy (e.g. Owen et al 1999).
\nocite{owe99}

Very massive stars emit photons with high enough energy to ionise their
surrounding gas, leading to a theoretical correlation \cite{bar97} between
the [OII]~3727 emission line luminosity and the mean star formation rate
over the past $10^7$ years (the approximate lifetime of such massive
stars):

\begin{equation}
\label{OIIstar}
\frac{\dot{M}_*}{M_{\odot}{\rm yr}^{-1}} \approx 6.3
\frac{L[OII]}{10^{41}\rm{erg\,s}^{-1}}.
\end{equation}

\noindent In practice, however, use of the [OII]~3727 emission line as an
estimator of star formation is fraught with problems (e.g. Hopkins et al
2001, Adelberger \& Steidel 2000, Kennicutt 1998 and references therein),
the most important of which is dust extinction which causes the star
formation rate to be underestimated. These problems can, however, be
mostly ignored if the level of accuracy required is only that to separate
radio sources into starbursting galaxies or AGN. The above conversion is
therefore adopted, but with an allowance for optical extinction. Assuming
that the average star formation rate over the past $10^8$ years (sampled
by the radio emission due to the synchrotron lifetime) is the same as that
over the past $10^7$ years (sampled by the [OII] line emission), then
equations~\ref{radstar} and~\ref{OIIstar} give a crude relationship
between the observed 5\,GHz radio flux density and the observed [OII]~3727
emission line flux for a star forming galaxy at redshift 0.83:

\begin{equation}
\label{radOII}
\frac{S_{\rm 5 GHz}}{\mu\rm{Jy}} \approx 4.4 \frac{f(\rm{[OII]~3727})}
{10^{-16}{\rm erg\,cm}^{-2}{\rm s}^{-1}} \frac{1}{T_{\rm [OII]}}
\end{equation} 

\noindent where $T_{\rm [OII]}$ is the fraction of light transmitted (not
extincted) at 3727\AA. The scatter in both of the line luminosity to star
formation rate and radio luminosity to star formation rate relations, and
the different star formation timescales that the two relations sample,
will provide further scatter around this relation.

No comparable relation is known between the [OII]~3727 and radio
luminosities for AGN at these radio luminosities, although the two
properties are well correlated over several orders of magnitude, albeit
with up to an order of magnitude of scatter, for more powerful radio
galaxies (see Figure~3 of Willott \etal\ 1999)\nocite{wil99b}; if this
relation can be extrapolated down in luminosity then, converting the
151\,MHz radio luminosity to a 5\,GHz radio luminosity assuming a spectral
index of 0.8:

\begin{equation}
\label{radOIIAGN}
\frac{S_{\rm 5 GHz}}{\mu\rm{Jy}} \approx 1.7 \times 10^4\left[\frac
{f(\rm{[OII]~3727})} {10^{-16}{\rm erg\,cm}^{-2}{\rm s}^{-1}}\right]^{1.45} 
\end{equation} 
 
\noindent It is not clear whether such an extrapolation is valid; for
example, Gruppioni et~al \shortcite{gru99} find that a significant
proportion of sub-mJy radio sources show only absorption lines and no
emission lines. Irrespective of the precise details, however, the critical
point is that for a given radio flux density the expected emission line
flux if the radio emission is produced by an AGN is at least 100--1000
times lower than if it is associated with star formation. A radio source
with $S_{\rm 5 GHz} \sim 100\mu$Jy will only have detectable [OII]~3727
line emission if at least some of the radio emission is due to star
formation.

\subsubsection{The nature of the cluster radio sources}

\begin{figure}
\centerline{
\psfig{file=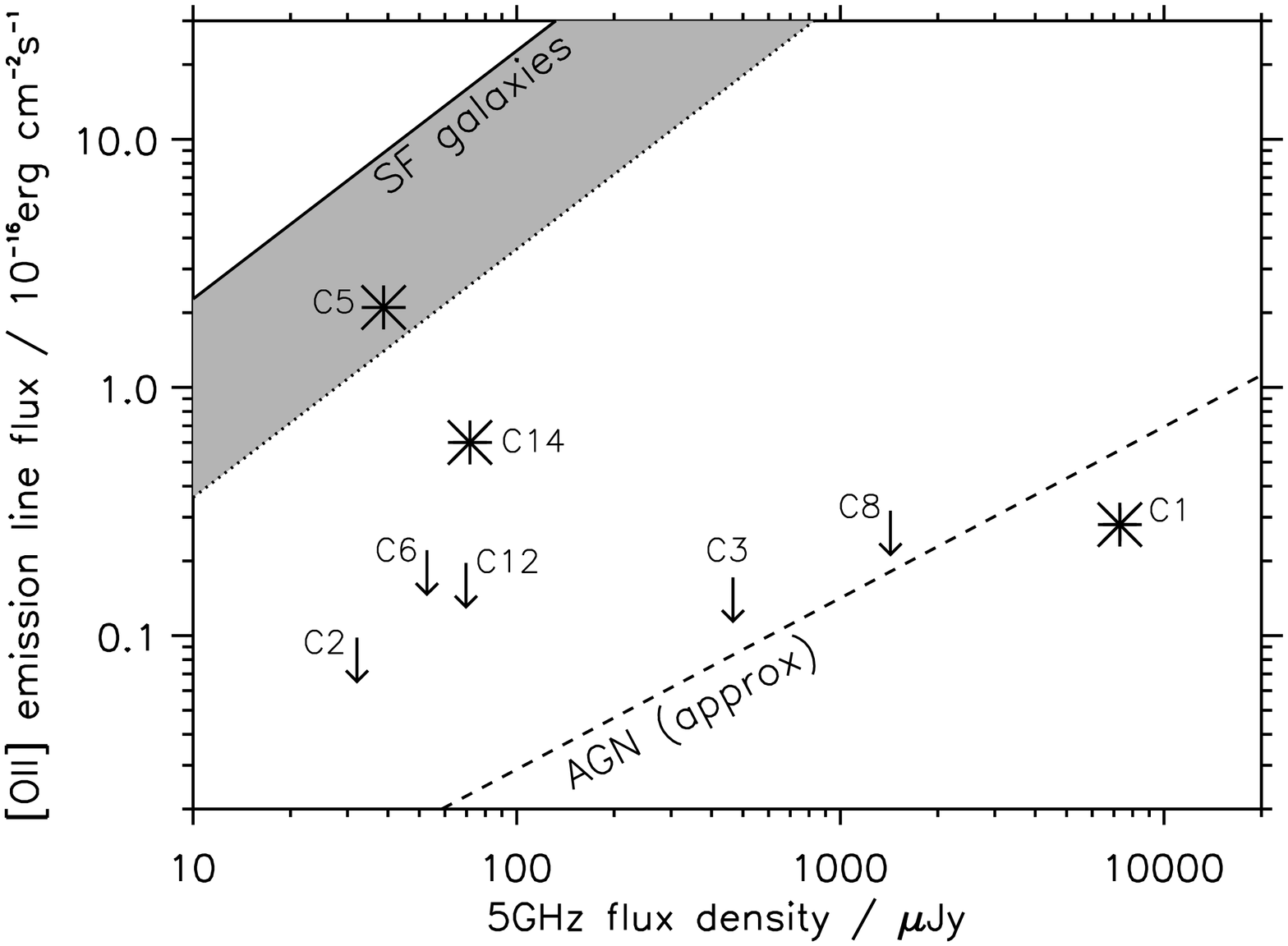,angle=90,width=8cm,clip=}
}
\caption{\label{emisrad} The distribution of [OII]~3727 line fluxes versus
radio flux densities for the cluster radio sources. The relation expected
for star forming galaxies (Barbaro \& Poggianti 1997) is indicated by the
solid line, and the shaded region represents the effect of up to two
magnitudes of dust extinction. The dashed line shows an extrapolation of
the Willott et~al (1999) relation for powerful radio galaxies; its
applicability at these lower radio luminosities is uncertain, but the key
point is that the for a given radio luminosity, the emission line
luminosity is 2--3 orders of magnitude lower for AGN than star-forming
galaxies.}
\end{figure}

Only three of the eight cluster radio sources (MS1054$-$C1, C5, C14) are
emission line galaxies (defined as [OII] equivalent width $> 5$\AA; van
Dokkum \etal\ 2000)\nocite{dok00}: MS1054$-$C1 has an [OII]~3727
equivalent width of $5.9 \pm 1.2$\AA, MS1054$-$C5 has $14.7 \pm 2.2$\AA\
and MS1054$-$C14 has $39.0 \pm 6.8$\AA. Using the continuum magnitudes and
colours from the HST observations to provide continuum calibration of the
spectra, the [OII]~3727 fluxes of these three sources are compared to
their radio fluxes in Figure~\ref{emisrad}. The other five radio galaxies
are similarly plotted, taking upper limits to the [OII]~3727 equivalent
width of 5\AA. Also indicated on the figure is the relation for starburst
galaxies derived in Equation~\ref{radOII}, with the shaded region
representing the effect of up to two magnitudes of optical extinction. The
extrapolated relation for AGN is also indicated, although once again it
must be stressed that this relation is only indicative.

It is clear from Figure~\ref{emisrad} that MS1054$-$C1 is an AGN, as would
be expected from its extreme radio luminosity. Equally, MS1054$-$C5 lies
well within the region of the plot predicted for star forming galaxies.
Its radio flux density corresponds to a star formation rate of nearly
$100\, M_{\odot}$yr$^{-1}$; its morphology and colour are consistent with
a high current star--formation rate.  MS1054$-$C14 lies between the AGN
and starburst regions, and so the origin of its radio emission is less
clear. This is likely either a heavily extincted starburst or a
combination of AGN and star formation. The other five cluster galaxies all
have upper limits to their emission line luminosity which place them well
below the star formation region. The spectra of all five of these galaxies
appear as typical elliptical galaxies. In two cases an extended radio
source has been produced, so these must be AGN; for the other three,
although a heavily obscured nuclear starburst cannot be excluded from
these sources, similar low luminosity radio sources in red ellipticals are
found in low redshift clusters and are seen to be weak AGN
\cite{owe99}. We conclude that these 5 sources are all likely to be weak
AGN.

Summarising, of the four radio sources associated with isolated galaxies,
2 are AGN, 1 is a starburst, and 1 is unclear. This is consistent with the
$50-60$\% starburst contribution at these luminosities which has been both
predicted on the basis of radio source count modelling (e.g. Wall and
Jackson 1997, Hopkins \etal\ 2000)\nocite{wal97,hop00}, and observed in
deep observations of the Hubble Deep Field \cite{ric98}. However, all four
of the radio sources associated with galaxy pairs appear to be AGN, and if
confirmed then the overall AGN fraction amongst the cluster sources would
be $\gta 75$\%, inconsistent with field observations at these flux density
levels. The most likely explanation for this is that the isolated cluster
radio sources discussed above are essentially an extension of the field
population, but that the galaxy pair radio sources represent a different
population: interactions between early--type galaxies, inducing weak AGN
activity, are unlikely to occur at a high rate in the field simply because
of the sparsity of early--types, and so this population may be fairly
unique to high redshift cluster environments.

\subsubsection{Cluster emission line galaxies}

None of the sources that van Dokkum \etal\ \shortcite{dok00} classified as
emission line galaxies (many of which also fall within the merger or blue
galaxy categories) are detected above the $2 \sigma$ level in the radio
map apart from the three sources discussed above (C1, C5 and C14). The
most emission line luminous cluster galaxy, 1801, has $f$([OII]) $\approx
3 \times 10^{-16}$erg\,cm$^{-2}$s$^{-1}$, predicting a $15\mu$Jy
(3$\sigma$) radio flux density, although this is not detected in the radio
maps. The mean radio emission of the emission line galaxies, $0.9
\pm 1.9 \mu$Jy, corresponds to an upper limit of about $5
M_{\odot}$yr$^{-1}$ which is comparable to the mean star formation rate of
these galaxies determined using Equation~\ref{OIIstar}.

It is clear from these observations that radio surveys directly seek out
both AGN and star--forming galaxies. A comparison with optical techniques
demonstrates that at the current depth these observations are picking out
the extreme end of the optically selected star--forming galaxies together
with a population of low luminosity active galaxies that are not found
using optical techniques. Still deeper observations will be required if
typical star--forming galaxies are to be observed.

\subsection{Cluster radio source luminosity distribution}
\label{lumdist}

Although the number of confirmed cluster radio sources in MS1054$-$03 is
very small, it is still instructive to perform some simple comparisons
with cluster radio populations at low redshift. Ledlow \& Owen
\shortcite{led96} have determined the univariate and bivariate luminosity
functions of radio sources in a sample of low redshift ($z < 0.09$)
clusters, from which can be determined the percentage of all elliptical
galaxies brighter than absolute R--magnitude $M_{\rm R} = -20.5$ which
host radio sources more luminous than a given radio luminosity. For a
galaxy with the colour of a typical early--type galaxy at $z=0.83$, and
taking into account also the $R$--magnitude change due to passive
evolution of the stellar populations, this optical magnitude limit
corresponds to an apparent magnitude through the HST F814W filter of $I
\sim 21.6$.

Within 1\,Mpc (168 arcsec) of the centre of MS1054$-$03 there are 23
early--type galaxies (no separation is made here between ellipticals and
S0 morphologies, and 3 of the 6 galaxies classified as ``S0/a'', ie
uncertain S0 or Sa, are included) with magnitudes $I < 21.6$ that are
confirmed to be members of the cluster; the spectroscopic observations are
currently over 80\% complete to this magnitude limit. Of these cluster
early--type galaxies, 5 show radio emission with a rest--frame 1.4\,GHz
luminosity above 23.1 W\,Hz$^{-1}$ (again, including only 1 of the 2 S0/a
classifications, and counting only one galaxy from each close pair). Given
that the spectroscopic observations were taken prior to the radio
observations, there is no reason to suspect that the spectroscopic targets
in any way favour those with radio emission, and hence the fraction $5/23
= 22\%$ should be a fair estimate of proportion of radio loud early--type
galaxies in the cluster. By comparison, Ledlow \& Owen
\shortcite{led96} found that only 5.8\% of elliptical galaxies (and hence
an even lower percentage of all early--type galaxies since very few of
their S0's were radio emitting) in low redshift clusters have radio
luminosities above this level. Clearly there is an excess of radio sources
in this high redshift cluster compared to low redshift clusters. However, 
this increase is only comparable to that found in the field: Hammer et~al
\shortcite{ham95} found that the corresponding fraction for $z \gta 0.7$
field ellipticals is of order a third.

Studying the Virgo cluster, Gavazzi \& Boselli \shortcite{gav99} have
shown that late type galaxies of all Hubble types are equally likely to
develop radio sources, with a probability proportional to the optical
luminosity of the galaxy. Only two of the radio sources are unambiguously
identified with late--type cluster galaxies (MS1054$-$C1 and C5), but it
is instructive that these are indeed two of the most luminous late--type
galaxies in the cluster (cf. van Dokkum et~al. 2000).

\begin{figure}
\centerline{
\psfig{file=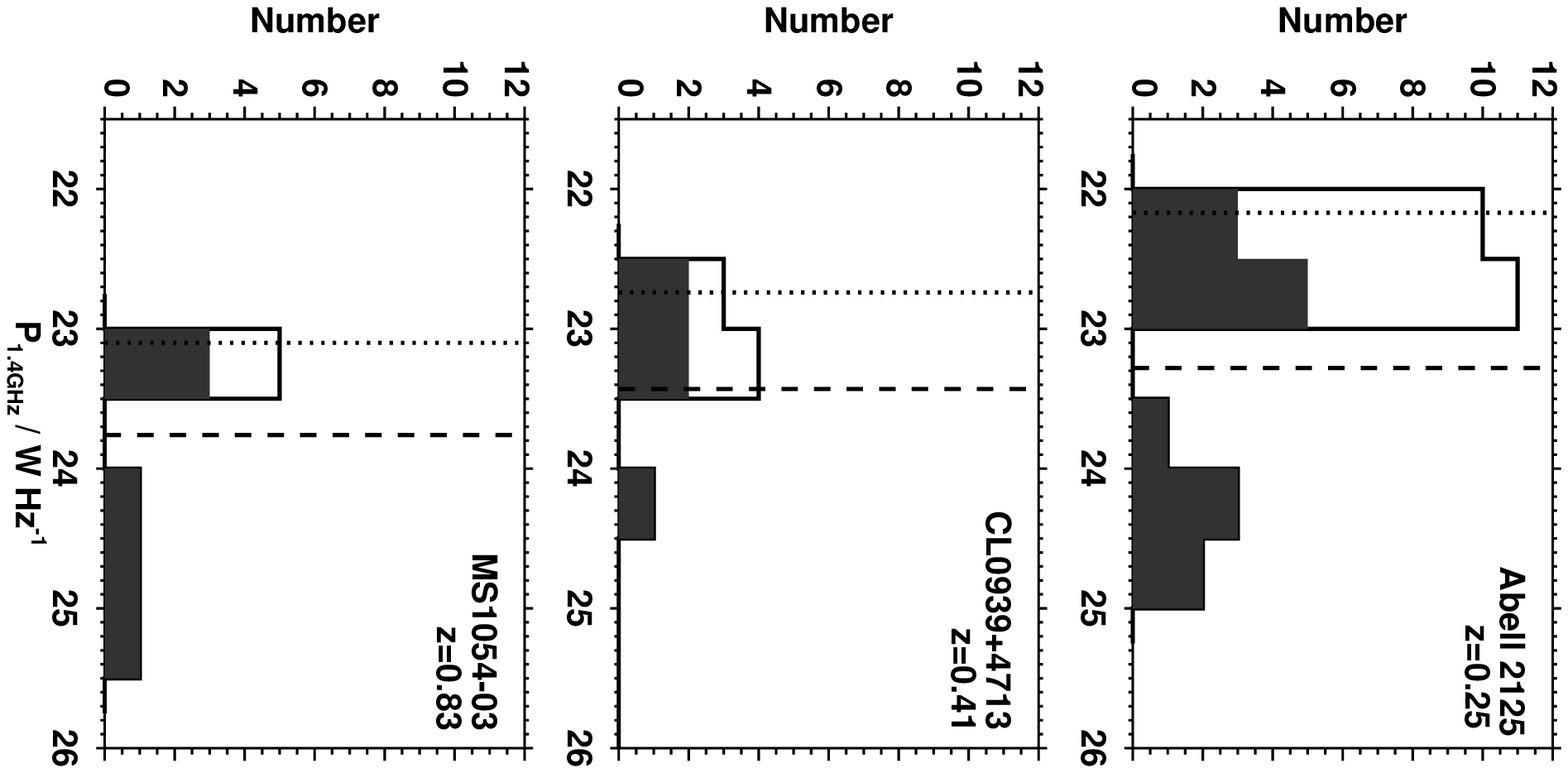,angle=90,width=8cm,clip=}
}
\caption{\label{nlumpop} {\it Top:} the distribution of radio luminosities
of radio sources associated with the $z=0.25$ cluster Abell 2125. The data
are from Dwarakanath \& Owen (1999), converted to the cosmology assumed in
this paper ($\Omega = 1$, $\Lambda = 0$, $H_{0}=
70$\,km\,s$^{-1}$Mpc$^{-1}$). The vertical dotted line represents the
radio luminosity detection limit of the sample, and the dashed line
indicates the centre of the break in the radio luminosity
distribution. The shaded region indicates the AGN population, and the
unshaded region comprises the starburst galaxies and unknown objects. The
sources classified as AGN by Owen et~al. (1999) had either AGN spectra or
were early--type galaxies with spectra showing no emission lines, thus
inconsistent with being star forming. {\it Middle:} the distribution of
radio source luminosities in the cluster CL0939+4713 ($z=0.41$) from Smail
et~al (1999), again converted to the above cosmology. The dotted line is
again the detection limit, and the dashed line shows the luminosity break
derived for A2125, scaled up in luminosity according to $(1+z)^3$
evolution. The AGN galaxies are similarly defined. {\it Bottom:} the
distribution of radio luminosities of sources whose host galaxies are
confirmed to be associated with MS1054$-$03 at $z=0.83$; the MS1054$-$03
data has been converted to a 1.4\,GHz rest--frame luminosity assuming a
spectral index of 0.8. The two vertical lines are as defined for
CL0939+4713.}
\end{figure}

As was discussed in the introduction, a deep radio survey of two $z \sim
0.25$ Abell clusters revealed a bimodality in the distribution of radio
source luminosities in that cluster (see Figure~\ref{nlumpop}a), with the
prevalence of the weaker radio source population providing a measure of
the state of activity of the cluster \cite{dwa99}. Those galaxies
classified as AGN by Owen et~al \shortcite{owe99} have been shaded in the
histogram: the more luminous radio source population is comprised entirely
of AGN, whilst the weaker population provides a mixture of AGN and
starbursts. Smail et~al. \shortcite{sma99b} studied the cluster CL0939+4713
at $z=0.41$, and their results are similarly plotted in
Figure~\ref{nlumpop}b. In Figure~\ref{nlumpop}c is reproduced the
histogram of radio luminosities of radio sources in the MS1054$-$03
cluster, converted from 5\,GHz to 1.4\,GHz rest--frame assuming a spectral
index of 0.8. The radio detection limit of each of the three clusters is 
indicated.

\nocite{owe99}

Whilst it cannot be over-stressed that these latter two clusters contain
very small numbers of confirmed cluster galaxies, it is of note that both
are consistent with the Abell 2125 distribution in the sense of being
plausibly bimodal, with the more luminous population consisting entirely
of AGN and the less luminous population being both AGN and starburst
galaxies. Further the break in luminosity between the two populations may
evolve to higher luminosities as we move to higher redshifts: certainly
the starburst galaxies detected in MS1054$-$03 are of higher radio
luminosity than any detected in the low redshift cluster. There are far
too few cluster radio sources to attempt to quantify whether, and how
quickly, any such evolution occurs. However, it is intriguing that the
$(1+z)^{\approx 3}$ luminosity evolution which matches the evolution of
the luminosity functions of many different types of active galaxies (e.g.\
radio sources -- Dunlop \& Peacock 1993\nocite{dun93}; optically--selected
quasars -- Maloney \& Petrosian 1999\nocite{mal99}; X-ray sources -- Boyle
\etal\ 1993\nocite{boy93}) fits the observations rather well.

\section{Conclusions}
\label{concs}

Deep radio imaging of the $z=0.83$ cluster MS1054$-$03 has revealed a
population of cluster radio sources. Eight radio sources are associated
with galaxies that are already spectroscopically confirmed cluster
members. This corresponds to approximately 22\% of early--type confirmed
cluster galaxies with $M_R < -20.5$ within 1\,Mpc of the cluster centre
having rest--frame 1.4\,GHz luminosity above 23.1 W\,Hz$^{-1}$. The
cluster radio sources fall into two distinct categories. 4 sources are
hosted by isolated galaxies which have a range of morphologies from
ellipticals to Sc galaxies; the radio emission appears to be associated
with a starburst in at least one case, and AGN activity in two. The other
four cluster radio sources are associated with close pairs of galaxies,
possibly interacting, although not all bound systems. 7 of the 8 galaxies
comprising these pairs are of type S0/a or earlier, and the radio emission
from these galaxy pairs is almost certainly nuclear in origin.  This $\gta
75$\% proportion of AGN in the cluster is in contrast to the field, where
$\sim 50-60$\% of radio sources at this flux density level are associated
with starburst galaxies. Considering the isolated cluster radio galaxies
alone, these are consistent with the field distribution and may
essentially be an extension of it, with the radio sources associated with
galaxy pairs being a new AGN population, driven by early--type galaxy
interactions, and thus fairly unique to high redshift cluster
environments.

MS1054$-$03 is a cluster which contains a high proportion of on--going
mergers. However, although up to 50\% of the confirmed cluster radio
sources appear to be interaction driven, none are associated with the
confirmed merger events, and the upper limit to the mean radio luminosity
of these merger events is of order $10^{22}$\,W\,Hz$^{-1}$, comparable to
that of M82. It appears that galaxy--galaxy interactions may be more
efficient than direct mergers at inducing radio emission.

Although the sample of confirmed cluster radio sources is small (8), the
cluster radio luminosity function shows a hint of bimodality, mirroring
that observed in the low redshift cluster Abell 2125, with a possible
increase in the break luminosity with redshift.

Clearly to investigate these issues further, spectroscopic redshifts for
the remainder of the radio sources in the sample will be required,
together with deep radio surveys of a larger sample of distant clusters 
to improve the statistics and confirm the results suggested here.

\section*{Acknowledgements} 

The National Radio Astronomy Observatory is operated by Associated
Universities Inc., under co-operative agreement with the National Science
Foundation. PNB would like to thank the Royal Society for generous
financial support through its University Research Fellowship scheme. PGvD
acknowledges support by NASA through Hubble Fellowship grant
HF-01126.01-99A awarded by the Space Telescope Science Institute, which is
operated by the Association of Universities for Research in Astronomy,
Inc., for NASA under contract NAS 5-26555. We thank the referee for useful
comments. 

\label{lastpage}
\bibliography{pnb} 
\bibliographystyle{mn} 

\appendix
\label{append}

\section{The individual sources}
\label{individ}

In this appendix the nature of each individual source is discussed,
ordered by radial distance from the centre of the cluster, and radio
contour maps of the field of each source are displayed in
Figure~\ref{overfigs}.  For those radio sources which lie within sky area
of the HST mosaic, the radio sources are shown overlaid upon a greyscale
representation of the F814W ($I$) image. In each case the greyscale range
is exactly the same, so the relative brightness of the objects can be
compared to that of the brightest cluster galaxy, visible in the
MS1054$-$C1 field. Those radio sources which lie outside of the HST mosaic
but within the larger field of the FORS imaging are shown overlaid upon
the $V$ band image, the greyscale range of the latter being adjusted to be
at comparable levels to that of the F814W image greyscales. The remaining
9 sources which lie outside of the FORS fields are simply displayed as
radio contour plots.

Details of the individual sources are as follows:

\smallskip
\noindent{\bf MS1054--C1:} First detected in the shallower imaging of
Stocke \etal\ \shortcite{sto99b}, this bright radio source is associated
with a confirmed cluster Sc galaxy. This galaxy is projected close to the
cluster centre (the elliptical galaxy visible 6.4 arcsec east and 8.3
arcsec south of this galaxy is the central cluster galaxy) although offset
in velocity by over 2000\kms. Its high radio luminosity ($1.3 \times
10^{25}$\,W\,Hz$^{-1}$) indicates that the radio emission must be
associated with AGN activity, and the radio to emission line flux density
ratio confirms this (Figure~\ref{emisrad}).

\smallskip
\noindent{\bf MS1054--C2:} The most tenuous radio detection in the sample,
this radio emission arises from the vicinity of three optically detected
galaxies, two of which are confirmed cluster members projected less than
20\,kpc apart, and the third of which lies at higher redshift
($z=1.076$). It is unclear whether the emission is associated with just
one of the galaxies, or with more, perhaps from a combination of two
cluster galaxies due to an interaction.

\begin{figure*}
\centerline{
\psfig{file=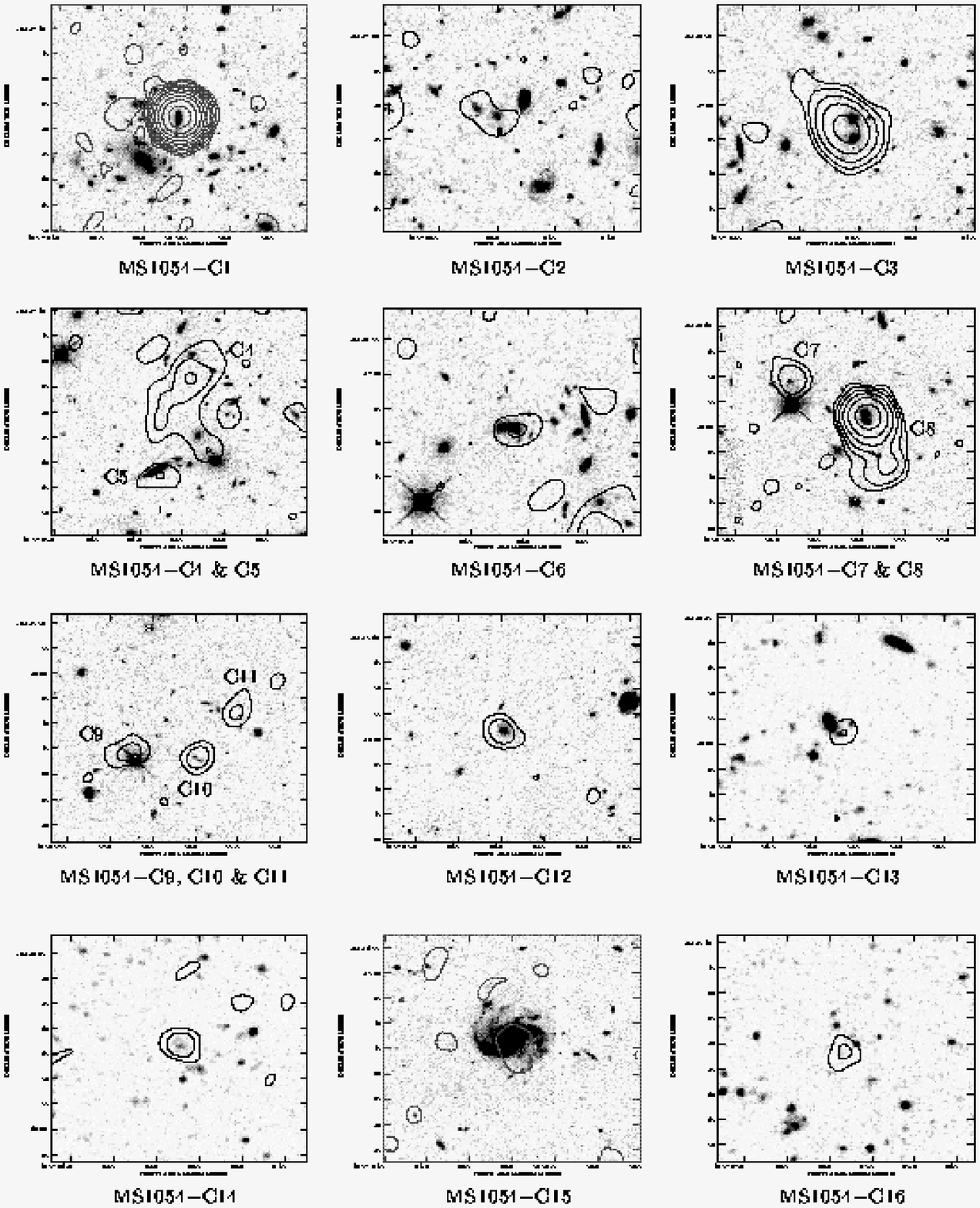,width=17cm,clip=}
}
\caption{\it see continuation for caption}
\end{figure*}

\addtocounter{figure}{-1}

\begin{figure*}
\centerline{
\psfig{file=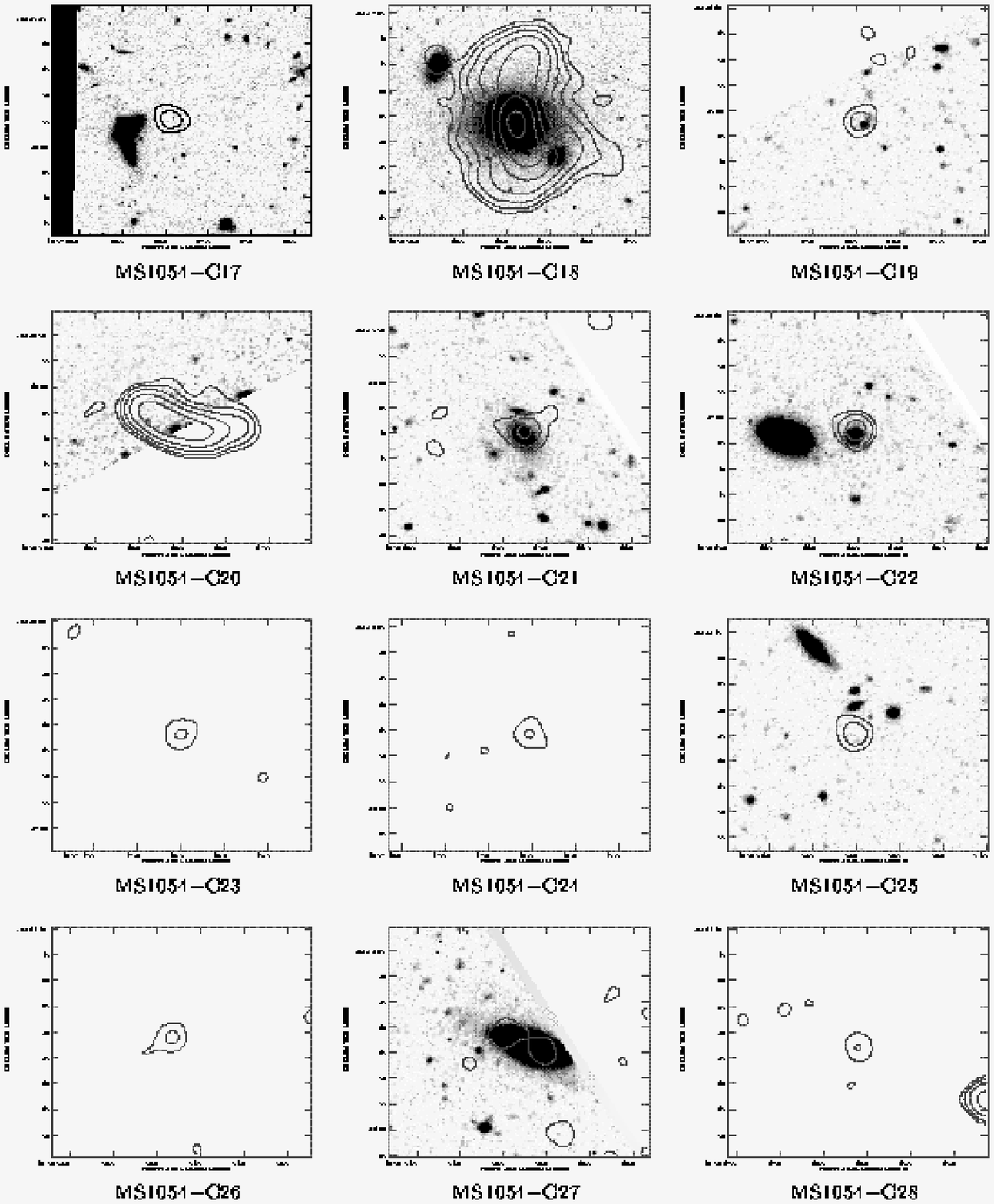,width=17cm,clip=}
}
\caption{{\bf cont.} {\it see continuation for caption}}
\end{figure*}

\addtocounter{figure}{-1}

\begin{figure*}
\centerline{
\psfig{file=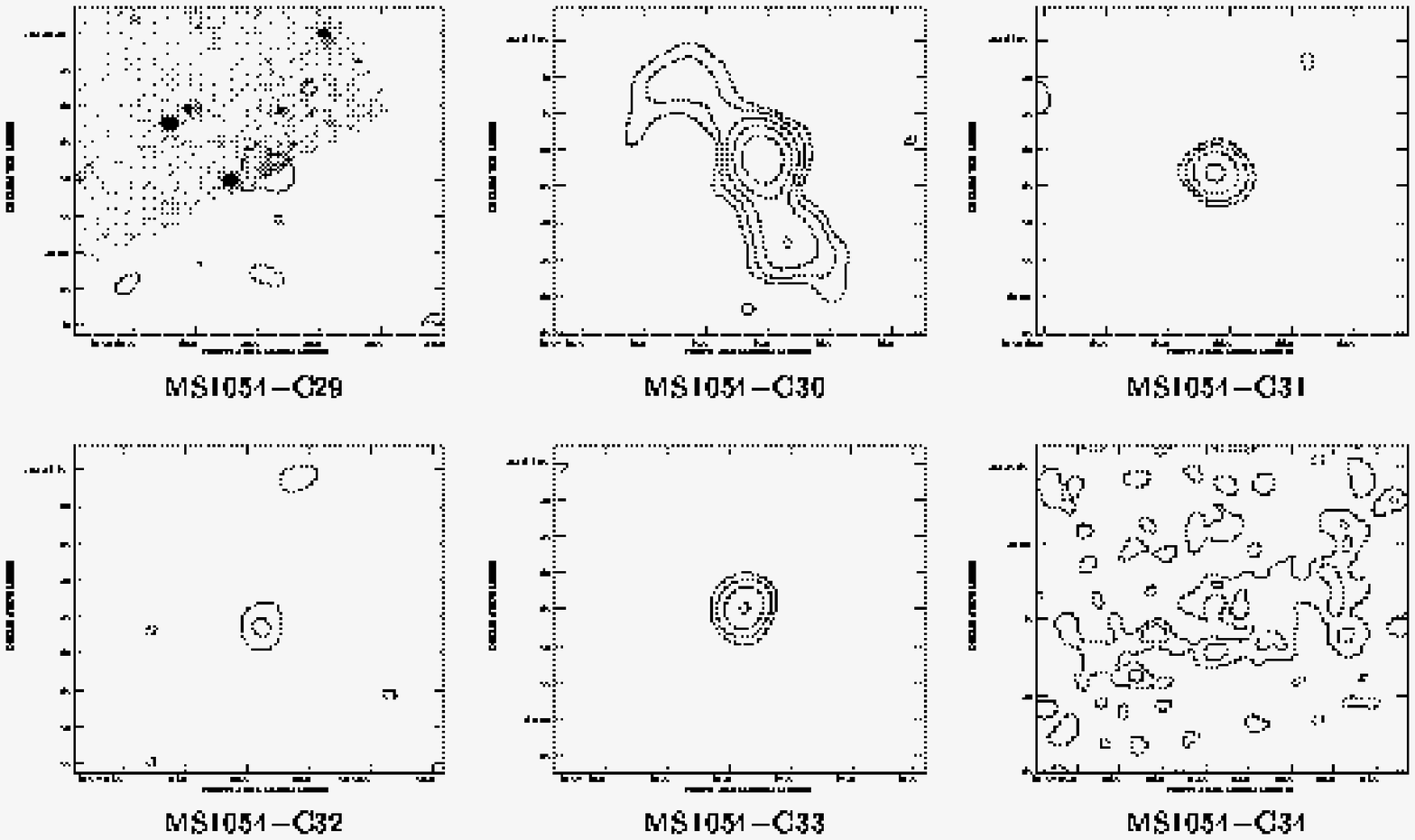,width=17cm,clip=}
}
\caption{\label{overfigs} Contour plots of the 34 radio sources detected
within 400 arcsec of the centre of the MS1054$-$03 cluster. Note that in
some cases, where the radio sources are projected close together, more
than one radio source is present in the plot. This is indicated by the
labelling on and underneath each figure. The contours are plotted at
($-$2, $-$1, 1, 2, 4, 8, 16, 32, 64, 128, 256, 512) $\times 16\mu$Jy
(3$\sigma$), except for MS1054$-$C34 where the contours at 8.0$\mu$Jy
(1.5$\sigma$) have also been added to better display the extended
structure, and MS1054$-$C4+5 where the first contour level was $13\mu$Jy
(2.5$\sigma$) to provide a clearer figure. Overlaid upon these contour
plots for the radio sources MS1054$-$C1, C2, C3, C4 \& C5, C6, C7 \& C8,
C9 \& C10 \& C11, C12, C15, C17 and C18 are greyscale images taken through
the F814W ($I$) filter of the HST; greyscale $V$ band images taken with
FORS on the VLT are overlaid for radio sources MS1054$-$C13, C14, C16,
C19, C20, C21 C22, C25, C27 and C29. Within each group (ie. the F814W and
the $V$ band images) the greyscale intensity range is held constant, and
has been adjusted to be comparable between the two groups. Thus the
brightness of the host radio galaxies can be visually compared to that of
the brightest cluster galaxy, which is the elliptical galaxy visible in
the field of MS1054$-$C1 at position 10 57 00.02, $-$3 37 36.0.}
\end{figure*}

\begin{figure*}
\centerline{
\psfig{file=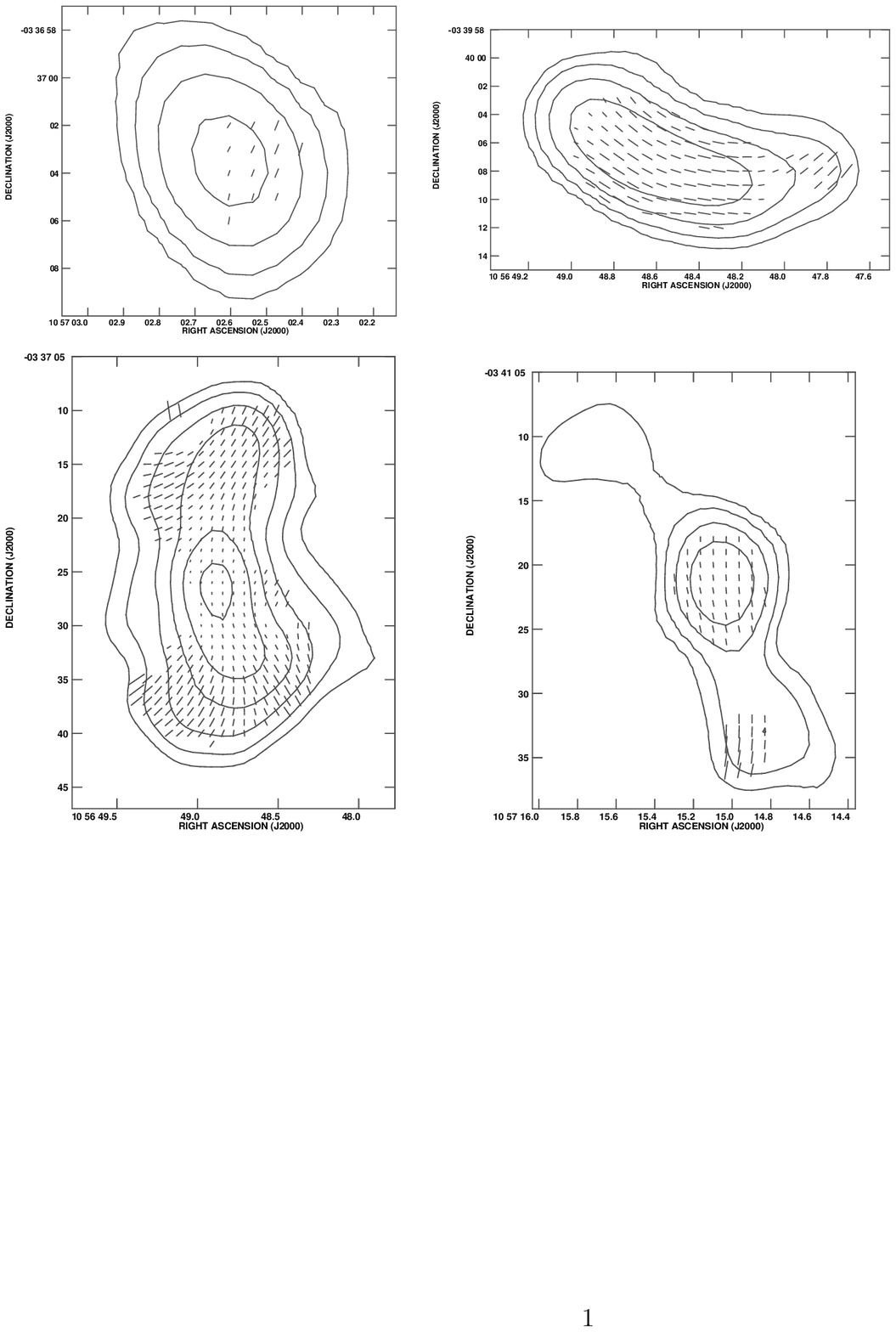,width=11cm,clip=}
}
\caption{\label{radpols} Contour plots with electric polarisation vectors
overlaid for the four radio sources which show polarised emission. The
contours are at ($-$1, 1, 2, 4, 8, 16, 32, 64) $\times$
32$\mu$Jy\,beam$^{-1}$, and a polarisation vector of length 1 arcsec
corresponds to 50\% polarisation. The sources are: upper left --
MS1054$-$C3; lower left -- MS1054$-$C18; upper right -- MS1054$-$C20;
lower right -- MS1054$-$C30.
}
\end{figure*}

\smallskip
\noindent{\bf MS1054--C3:} This relatively bright radio source was also
detected in the shallower observations of Stocke \etal\
\shortcite{sto99b}.  It lies between two confirmed cluster galaxies, not
especially consistent with the location of either (though note that the
position determined by Stocke et~al. places it closer to the southern
galaxy). The two galaxies may be interacting, separated by a projected
distance of only 17\,kpc, although the colours of both galaxies are red:
one galaxy lies on the `passive--evolution' colour--magnitude relation and
the other is about 0.2 magnitudes bluer than this
(Figure~\ref{colmagdiag}). The radio source is extended by of order the
separation of the two galaxies, and shows a marginal detection of
polarisation (see Figure~\ref{radpols}), indicating that the radio
emission is likely to have an AGN origin.

\smallskip
\noindent{\bf MS1054--C4:} This extended region of radio emission has no
obvious optical counterpart in the deep HST images. Its nature is unclear,
although its luminosity and radial offset from the cluster centre are both
consistent with it being a cluster relic radio source: these are diffuse
radio sources found occasionally in clusters, associated with no
particular parent galaxy (e.g. Giovannini et~al 1999 and references
therein).\nocite{gio99} Their origin is still uncertain.

\smallskip
\noindent{\bf MS1054--C5:} This region of weak radio emission appears to
be separate from the extended source MS1054$-$C4, although it cannot be
ruled out that these form part of the same source. Far more likely is that
this emission is associated with starbursting activity in the $z=0.8113$
cluster Sc galaxy projected only 1.2 arcsec away: as shown in
Section~\ref{emisstren}, the radio luminosity of the source is very close
to that which would be predicted on the basis of the emission line
luminosity of this Sc galaxy, assuming both to be associated with
on--going star formation.

\smallskip
\noindent{\bf MS1054--C6:} This radio emission is associated with a
confirmed cluster elliptical galaxy, whose isophotes overlap with a second
elliptical galaxy of comparable magnitude and colour (but unknown
redshift). Other fainter galaxies are observed close-by. If the second
elliptical galaxy is also a cluster member, then the radio emission is
likely to have been induced by galaxy--galaxy interaction.

\smallskip
\noindent{\bf MS1054--C7:} Although there is a small possibility that this
radio emission forms part of the extended radio source MS1054$-$C8, it
appears to be a separate radio source associated with a 22nd magnitude galaxy, 
whose redshift is currently unknown.

\smallskip
\noindent{\bf MS1054--C8:} This bright radio source was first detected by
Stocke \etal\ \shortcite{sto99b} and is associated with two galaxies both
of which are confirmed cluster members with a projected separation of
11\,kpc. These galaxies may be interacting, but do not show any features
that would cause them to have been classified as a merger by van Dokkum
et~al. \shortcite{dok99a}. The extended structure of this AGN is
consistent with a core--jet source. A core--dominated (ie. beamed) quasar
is unlikely since neither of the associated galaxies has magnitudes or
colours indicating a strong nuclear component.

\smallskip
\noindent{\bf MS1054--C9:} This unresolved weak radio source has no
obvious optical identification on the deep HST mosaic, although the
location of the source, on the diffraction spike of a nearby 19th
magnitude star, may have precluded any faint optical galaxy from being
identified on the HST image. This star provides the entry in
Table~\ref{radopttab}, but is unlikely be the true radio source host given
the radio--optical offset.

\smallskip
\noindent{\bf MS1054--C10:} A pair of very faint ($I \sim 23.5$)
blue galaxies is seen at the location of this radio source, although
redshifts are not available for either of them.

\smallskip
\noindent{\bf MS1054--C11:} The 23rd magnitude galaxy which lies within
half an arcsecond of the position of this weak radio source has no
measured redshift.

\smallskip
\noindent{\bf MS1054--C12:} This unresolved radio source is
unambiguously identified with an early type galaxy at the redshift of
the cluster.

\smallskip
\noindent{\bf MS1054--C13:} This radio source corresponds to a 22nd
magnitude Sab galaxy of unknown redshift. The brighter galaxy a few arcsec
to the north--east of the radio source has a redshift of $z=0.17$, and is
unlikely to be related.

\smallskip
\noindent{\bf MS1054--C14:} This radio source is associated with a diffuse
optical galaxy at the cluster redshift.

\smallskip
\noindent{\bf MS1054--C15:} This weak extended radio emission appears to
be associated with a bright spiral galaxy at redshift 0.25. The measured
flux density corresponds to a radio luminosity of $1.2 \times
10^{22}$\,W\,Hz$^{-1}$, similar to that of M82.

\smallskip
\noindent{\bf MS1054--C16:} No optical counterpart brighter than $I
\approx 26$ is seen associated with this unresolved radio source.

\smallskip
\noindent{\bf MS1054--C17:} At the location of this radio source there is
only a marginal detection of an extremely faint galaxy ($I \sim 26$).

\smallskip
\noindent{\bf MS1054--C18:} This luminous extended radio source, which has
an edge--darkened structure compatible with being an FR\,I type source
\cite{fan74} is associated with a low redshift ($z=0.182$) elliptical
galaxy which has a second galaxy projected 3 arcsec away to the NNE
(within the saturated region of the presented greyscale), almost certainly
undergoing an interaction. The radio source is quite strongly polarised
(see Figure~\ref{radpols}) with a well--ordered polarisation structure
along the source axis in the central regions but perpendicular to this at
the extremities.

\smallskip
\noindent{\bf MS1054--C19:} An optical galaxy lies within half an
arcsecond of the position of this possibly resolved radio source, but its
redshift is unknown.

\smallskip
\noindent{\bf MS1054--C20:} This extended source could have either FR\,I
or head--tail structure and shows strong polarisation throughout,
parallel to the extended radio structure (Figure~\ref{radpols}). It
lies at the very edge of the optical frame, but appears to have an
associated optical counterpart.

\smallskip
\noindent{\bf MS1054--C21:} A galaxy at redshift $z=0.374$ lies roughly
coincident with this weak unresolved radio source. The corresponding
rest--frame 5\,GHz radio luminosity is $4 \times 10^{22}$\,W\,Hz$^{-1}$,
consistent with the radio emission being associated with star formation
activity. 

\smallskip
\noindent{\bf MS1054--C22:} This unresolved radio source has a
relatively unambiguous optical counterpart which has not been observed 
spectroscopically.

\smallskip
\noindent{\bf MS1054--C23:} This unresolved radio source lies outside
the field of the optical images.

\smallskip
\noindent{\bf MS1054--C24:} This unresolved radio source lies outside
the field of the optical images.

\smallskip
\noindent{\bf MS1054--C25:} No optical counterpart is seen at the
location of this radio source.

\smallskip
\noindent{\bf MS1054--C26:} This unresolved radio source lies outside
the field of the optical images.

\smallskip
\noindent{\bf MS1054--C27:} A low redshift spiral galaxy is clearly the
source of this extended radio emission. Although the galaxy has not been
observed spectroscopically, its bright magnitude suggests a low redshift,
implying that the radio emission is likely to be associated with on-going
star formation.

\smallskip
\noindent{\bf MS1054--C28:} This unresolved radio source lies outside
the field of the optical images.

\smallskip
\noindent{\bf MS1054--C29:} This radio source is possibly resolved and
lies at the very edge of the optical image. An optical galaxy partially
seen at the edge of the image appears to be associated with it.

\smallskip
\noindent{\bf MS1054--C30:} Extending nearly half an arcminute on the
sky, this radio source is clearly brightest at its centre and probably
has FR\,I structure, although an FR\,II structure with a bright radio
core cannot be excluded. The central radio region is highly polarised,
with the electric field vectors lying parallel to the extended radio
structure. The radio source lies outside the field of the optical
images.

\smallskip
\noindent{\bf MS1054--C31:} This unresolved radio source lies outside
the field of the optical images.

\smallskip
\noindent{\bf MS1054--C32:} This unresolved radio source lies outside
the field of the optical images.

\smallskip
\noindent{\bf MS1054--C33:} This unresolved radio source lies outside
the field of the optical images.

\smallskip
\noindent{\bf MS1054--C34:} A large diffuse area of radio emission,
extending an arcminute or more, is detected at this location. It lies
outside of the optical frames, and no luminous nearby galaxies are seen in
the digitized sky catalogues at this location, so its nature remains
unclear. One possibility again is that this is associated with a cluster
relic radio source; although it is found much further from the centre of
the cluster (2.2\,Mpc) than is typical for these sources, such large radii
relics have been observed before (e.g. in Abell 2345; Giovannini
et~al. 1999).

\end{document}